\def\VEL{\:{\rm km\:s^{-1}}}

\documentclass[11pt,a4paper]{article}

\newcommand{\MSOL}{\mbox{$\:M_{\odot}$}}

\usepackage{ascmac}
\usepackage{amsmath}
\usepackage{amssymb}
\usepackage{bm}
\usepackage[footnotesize,bf]{caption}
\usepackage{fullpage}
\usepackage{color}
\usepackage{float}
\usepackage{graphicx}
\usepackage[]{natbib}
\usepackage{subfigure}
\usepackage{txfonts}
\usepackage{threeparttable}
\usepackage{wrapfig}
\usepackage[]{multicol}
\usepackage{wrapfig}
\usepackage{authblk}

\usepackage{floatflt}

\title{\large Complete list of the ASTRO-H Science Working Group}
\date{\vspace{-0.5cm}}
\newcommand{\MakeWhitePaperTitle}{
	\begin{center}
		\begin{figure}
			\vspace{1cm}
			\begin{center}
				\includegraphics[width=0.2\hsize]{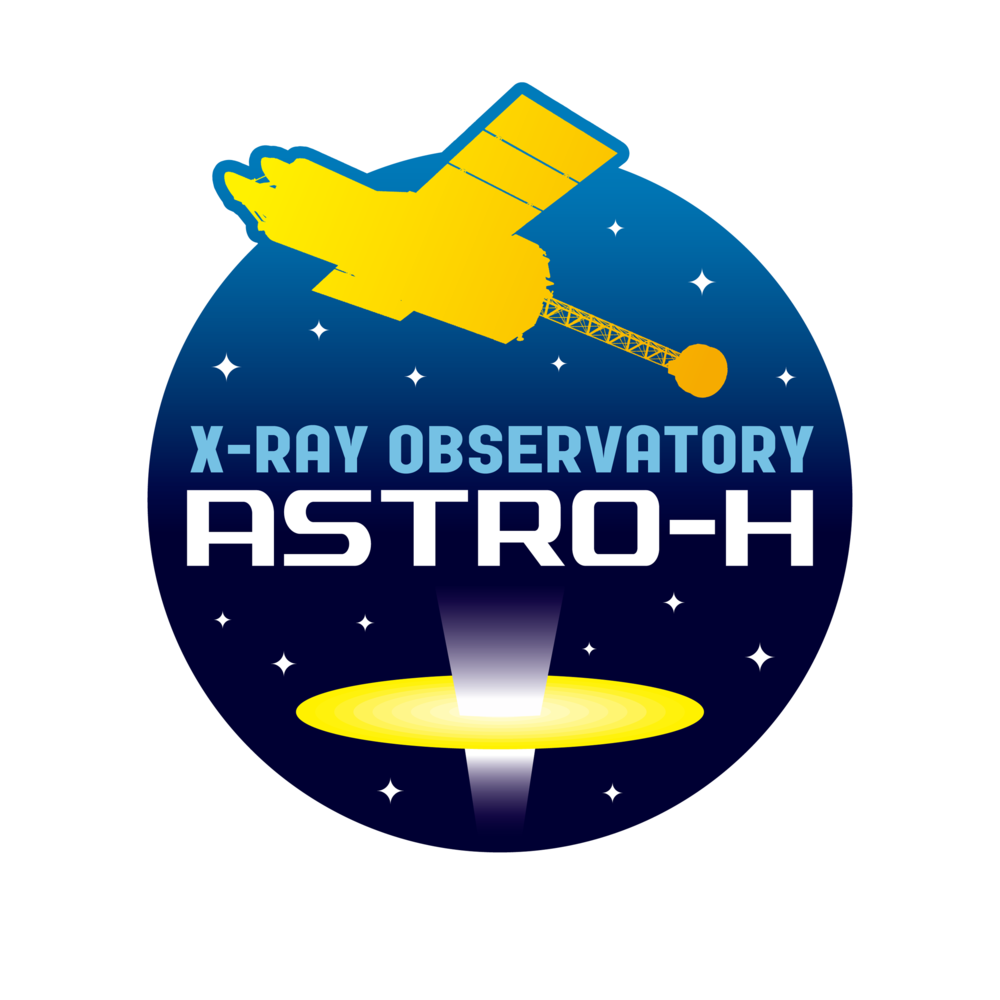}
			\end{center}
		\end{figure}
		\vspace{1cm}
		{\LARGE
		ASTRO-H Space X-ray Observatory\\
		White Paper\\
		}
		\vspace{5mm}
		{\large
		\WhitePaperTitle\\
		}
		\vspace{1cm}
		{
		\WhitePaperAuthors\\
		on behalf of the ASTRO-H Science Working Group
		}
	\end{center}
}

\usepackage{authblk}
\author[a]{Tadayuki~Takahashi}
\author[a]{Kazuhisa~Mitsuda}
\author[b]{Richard~Kelley}
\author[c]{Felix~Aharonian}
\author[d]{Hiroki~Akamatsu}
\author[e]{Fumie~Akimoto}
\author[f]{Steve~Allen}
\author[g]{Naohisa~Anabuki}
\author[b]{Lorella~Angelini}
\author[h]{Keith~Arnaud}
% ...
\affil[a]{Institute of Space and Astronautical Science (ISAS), Japan Aerospace Exploration Agency (JAXA), Kanagawa 252-5210, Japan}
\affil[b]{NASA/Goddard Space Flight Center, MD 20771, USA}
\affil[c]{Astronomy and Astrophysics Section, Dublin Institute for Advanced Studies, Dublin 2, Ireland}
\affil[d]{SRON Netherlands Institute for Space Research, Utrecht, The Netherlands}
\affil[e]{Department of Physics, Nagoya University, Aichi 338-8570, Japan}
\affil[f]{Kavli Institute for Particle Astrophysics and Cosmology, Stanford University, CA 94305, USA}
\affil[g]{Department of Earth and Space Science, Osaka University, Osaka 560-0043, Japan}
\affil[h]{Department of Astronomy, University of Maryland, MD 20742, USA}
% ...

\begin{document}

\newcommand{\WhitePaperTitle}{Older Supernova Remnants and Pulsar Wind Nebulae}
\newcommand{\WhitePaperAuthors}{
	K.~S.~Long~(STScI), A.~Bamba~(Aoyama~Gakuin~University), 
	F.~Aharonian~(DIAS~\&~MPI-K), ~
	A.~Foster~(Harvard-Smithsonian~Center~for~Astrophysics),
	S.~Funk~(Stanford~University), 
	J.~Hiraga~(University~of~Tokyo), J.~Hughes~(Rutgers~University),
	M.~Ishida~(JAXA), S.~Katsuda~(JAXA), ~H.~Matsumoto~(Nagoya~University), 
	K.~Mori~(Miyazaki~University), 
	H.~Nakajima~(Osaka~University), ~T.~Nakamori~(Yamagata~University),
	M.~Ozaki~(JAXA),
	S.~Safi-Harb~(University~of~Manitoba), 
	M.~Sawada~(Aoyama~Gakuin~University), T.~Tamagawa~(RIKEN), 
	K.~Tamura~(Nagoya~University),
	T.~Tanaka~(Kyoto~University), H.~Tsunemi~(Osaka~University),
	H.~Uchida~(Kyoto~University), Y.~Uchiyama~(Rikkyo~University),
	and
	S.~Yamauchi~(Nara~Womens~University)
}
\MakeWhitePaperTitle

\begin{abstract}
Most supernova remnants (SNRs) are old, in the sense that their structure has been profoundly modified by their interaction with the surrounding interstellar medium (ISM).  Old SNRs are very heterogenous in terms of their appearance, reflecting differences in their evolutionary state, the environments in which SNe explode and in the explosion products.  Some old SNRs are seen primarily as a result of a strong shock wave interacting with the ISM. Others, the so-called mixed-morphology SNRs, show central concentrations of emission, which may still show evidence of emission from the ejecta.  Yet others, the pulsar wind nebulae (PWNe), are seen primarily as a result of emission powered by a pulsar; these SNRs often lack the detectable thermal emission from the primary shock.  The underlying goal in all studies of old SNRs is to understand these differences, in terms of the SNe that created them, the nature of the ISM into which they are expanding, and the fundamental physical processes that govern their evolution.  Here we identify three areas of study where {\it ASTRO-H} can make important contributions.  These are constraining abundances and physical processes  in mature limb-brightened SNRs,  understanding the puzzling nature of mixed-morphology SNRs, and  exploring the nature of PWNe.   
The Soft X-ray Spectrometer (SXS) on-board {\it ASTRO-H} will, as a result of its high spectral resolution, be the primary tool for addressing problems associated with old SNRs, supported by hard X-ray observations with the Hard X-ray Imager (HXI)  to obtain broad band X-ray coverage.
\end{abstract}

\maketitle
\clearpage

\tableofcontents
\clearpage

Supernova remnants (SNRs) represent a crucial phase in the life cycle of stars and galaxies, a phase during which the energy of a SN explosion is transferred to the interstellar medium (ISM) and the material generated in stars is returned to the ISM to create the next generation of stars. They also represent laboratories for understanding the physics of both thermal and non-thermal plasmas with conditions that are far different from what can be produced in any earth-based laboratory.
 
Most SNRs are old in the sense that they have swept up much more material than that was ejected by the exploding star.   As a result, most, though not necessarily all, of the observed X-rays arise from the interaction of the shock with the ISM.  If the ISM were uniform, then all SNRs would evolve through a set of well-defined stages - a free expansion phase, an adiabatic phase, and a radiative phase before merging into the ISM. Similarly, if all SN explosions were identical, we would not find some SNRs with central objects and pulsar wind nebulae and others not.  Instead, SNRs are very heterogenous in appearance as a result of a complex  
 set of factors ranging from the nature of the SN explosion to the density and uniformity of the ISM into which they are expanding.   Ideally, we would like to reduce this apparent complexity into a small number of variables that are connected concretely to the SN progenitor and the environment.
 
 Young SNRs will be addressed in a separate white paper (Hughes et al.).  Here we focus on the older SNRs, which are dominated by shocked ISM material, and on Pulsar Wind Nebulae (PWNe), the synchrotron bubbles inflated by rapidly rotating neutron stars born in core-collapse explosions. Studies of these objects shed light on basic astrophysical questions related to shock and plasma physics, the evolution of SNRs and interaction with the ISM, and the way neutron stars deposit energy into and interact with the surrounding medium.

Old SNRs can be divided into 3 subgroups: (1) Limb-brightened SNRs in which most of the observed X-ray emission arises just behind the primary shock front as it moves into the ISM, (2) Mixed-morphology SNRs in which most of the emission arises from a thermal plasma in the interior of the SNR, and (3) Pulsar Wind Nebulae,  in which most of the emission arises from radiation from energetic electrons produced by an active pulsar.

\section{Limb brightened SNRs and the physics of shocks\label{shocks}}

\subsection{Background and Previous Studies}

When a SN explodes, a strong shock is sent out into the ISM.  
This shock heats the material it encounters to temperatures which 
depend on the speed of the shock and the fraction of the shock 
energy that is thermalized at the shock front.  In older SNRs, shock speeds are typically $100-500$~km~s$^{-1}$ and 
as a result, the plasma behind the shock typically has a temperature in the range $10^6$ to $10^7$~K.   The X-ray radiation from this thermal plasma
is dominated by emission lines.  Studies of these emission lines provide crucial information about plasma conditions behind the shock, the abundance patterns in the post-shock gas, and
atomic processes that are important in shocked plasmas.

The first imaging surveys of old SNRs were carried out using {\it Einstein} and {\it ROSAT}.   The data were typically analyzed in terms of 1-dimensional Sedov models in order to infer basic properties of the SNRs, such as their approximate age and evolutionary state and the density of the ISM into which they were expanding.  With the advent of  CCDs, it became possible to observe individual emission lines in these SNRs, particularly at energies greater than 1 keV.

\subsection{Prospects \& Strategy}

The rims of old SNRs are ideal sites to study dynamics of the cloud-shock 
interactions and of the atomic processes associated with relatively slow shocks, because line broadening is free from not only thermal 
Doppler broadening but also the bulk motions of the expanding shock.  For shocks associated with 
old SNRs (the velocities are typically $100--500~km~s^{-1}$), 
it is observationally known that electrons and ions are equilibrated 
immediately behind the shock, making the ion temperature 1\,keV or lower.  Thus, expected thermal Doppler broadening is 
$\lesssim0.1$\,eV, even though the shock is collisionless 
\citep[e.g.,][]{Ghavamian2007}.  Also, the bulk motion due to shell expansion is not of concern
in the rim regions of largely extended SNRs.  Thus, old 
SNRs provide clean laboratories for the study of plasma dynamics 
due to cloud-shock interactions.  

There are a number of problems associated with old SNRs with limb-brightened morphologies that can be addressed with {\it ASTRO-H}:

First, CCD measurements of the post-shock gas of the rims of old SNRs yield metal abundances significantly lower than average Galactic ISM values.  This is the case for the Cygnus Loop 
\citep[e.g.,][and references therein]{miyata94,Katsuda2008,Uchida2009}, 
Puppis~A \citep{Hwang2008}, RCW86 \citep{Vink1997,Yamaguchi2008} 
and the Vela SNR \citep{Miceli2008}.   Why this should be so is unclear.  One possibility is that the shocks at the rim of old SNRs are moving into material polluted by mass loss from the pre-SN star; another is that metals are bound up in dust grains.  Yet a third possibility is that atomic processes such as charge-exchange (CX) emission \citep{katsuda11} and resonance line 
scattering \citep{Miyata2008}, which are not taken into account in the models, affect the inferred abundances.   A final possibility is that a inaccuracies in the models and in instrumental characteristics are introducing systematic errors in the analysis. 
Analysis of CCD-based data is strongly dependent both on models and on a 
precise understanding of instrumental parameters, because with CCD spectral resolution of $\sim$100 eV, the lines used to determine abundances are often blended and not cleanly separable from the continuum,
especially below 1.2 keV.   Although similar problems affect observations of many other types of objects,  the  problem is especially severe for old SNRs, since their X-ray spectra are soft, with characteristic temperatures less than 1 keV.   Many lines of N, O, Ne and Fe contribute to the spectrum between 300~eV and 1200~eV.   With an energy resolution better than 7~eV, 
the {\it ASTRO-H} SXS  will dramatically reduce the problems due to limited spectral resolution, and allow us to pinpoint the real cause of the low abundances reported previously. 

Second, although old SNRs with limb-brightened X-ray morphologies are dominated by emission from the ISM, many, including the Cygnus Loop, Vela, RCW86 and Puppis A, show evidence of ejecta in a hotter component that is more centrally peaked than the emission from the ISM shock.  Although it has been established that RCW86 is most likely the result of a Type Ia explosion, whereas the Cygnus Loop, the Vela SNR, and Puppis A are remnants of core-collapse explosions, Studies of this material using the {\it ASTRO-H} SXS are crucial in order to determine the mass and energetics of the SN explosion that produced each of these remnants. 

\subsection{Targets \& Feasibility}

There are a large number of shell-like SNRs that could be observed with {\it ASTRO-H}.  The two prototypical shell-like SNRs for X-ray studies are the Cygnus Loop and the Vela SNR.  They have become the prototypes primarily because they are nearby and bright, which meant they were easily resolvable with the first generation imaging telescopes.  They remain the best objects for studies of older SNRs for roughly the same reasons with {\it ASTRO-H}, and we discuss them in more detail below.  

Other SNRs of interest include RCW86, which appears to be an 
example of a Type Ia explosion within a wind-blown bubble \citep{Vink1997}; Puppis A, which, based optical spectra and the presence of a central compact remnant, is the result of a core-collapse SN; plus  
a number of SNRs in the Magellanic Clouds.  All have high surface brightness and therefore good spectra can be obtained with {\it ASTRO-H} SXS in reasonably short exposures.


\begin{figure}[htb]
\begin{center}
\includegraphics[scale=0.4,angle=0]{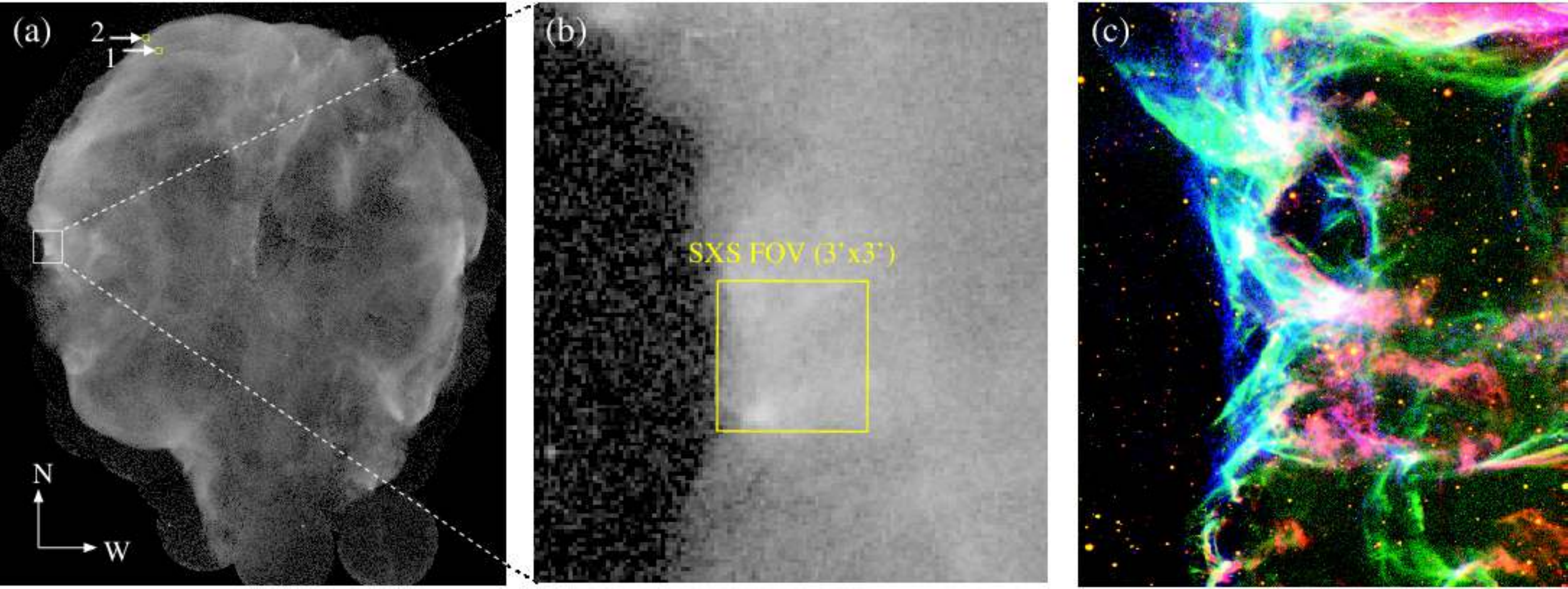}
\caption{Left: X-ray view of the entire Cygnus Loop taken by the ROSAT 
HRI \citep{Levenson1997}.
The so-called XA region is located within the white box.  SXS 
simulations for the arrow-indicated yellow box regions are presented 
in Figure~\ref{fig:NEcyg_spec}.     
Center: Close-up image of the white box region in the left panel.
The SXS FOV is shown as a yellow box.  
Right: Optical image covering the roughly same region as in the center 
panel \citep[the image is taken from][]{Danforth2001}.  Red, green, 
and blue correspond to H$\alpha$, [O III], and the FUV, respectively.\label{fig:XAcyg_image}
}
\end{center}
\end{figure}

\subsubsection{Cygnus Loop - A prototypical limb-brightened SNR}

The Cygnus Loop is one of the brightest soft X-ray sources.  It is limb brightened at X-ray and radio wavelengths, and most of the observed optical emission also arises from the shell.  About 3$^\circ$ in diameter, with a ``breakout'' region in the south, it lies at a distance of 540\,pc \citep{blair05}, implying it has a diameter of about 28 pc.  It is associated with the delicate optical filamentation of the Veil Nebula, which includes both Balmer-dominated and radiative shocks \cite[see, e.g.][]{levenson98}.   The Cygnus Loop is thought to be the result of a Type II SN explosion that took place about 10$^4$\,years ago (even though no stellar remnant has been found).  Measured shock velocities range from about 150 to 400 $\VEL$.   It is believed to be in an transition from the adiabatic to the radiative phase.   The current X-ray morphology is thought to be due to the encounter of primary shock wave with the wall of a cavity that was generated by the pre-SN star \citep{mccray79, levenson98}.  

The Cygnus Loop has a two component thermal X-ray spectrum \citep{tsunemi07}.  The first component, which appears to have sub-solar metal abundances and  a low temperatures ($kT_e\sim0.2$~keV), is dominant along the rim, and is associated with the primary shock moving into the ISM and/or cavity wall.   Since the shell is capturing the fresh ISM, it is in an ionizing phase \citep{miyata94}.  The ionization timescale is around 10$^{11}$~cm$^{-3}$~s in the shell region.  The second component, found in the interior of the SNR,  shows higher temperatures ($kT_e\sim0.7$~keV).  The second component is metal rich, with Si:H ratios a few times solar  \citep{uchida09},  implying that it arises at least in part from ejecta of the SN explosion.

\begin{figure}[h]
\includegraphics[width=16cm]{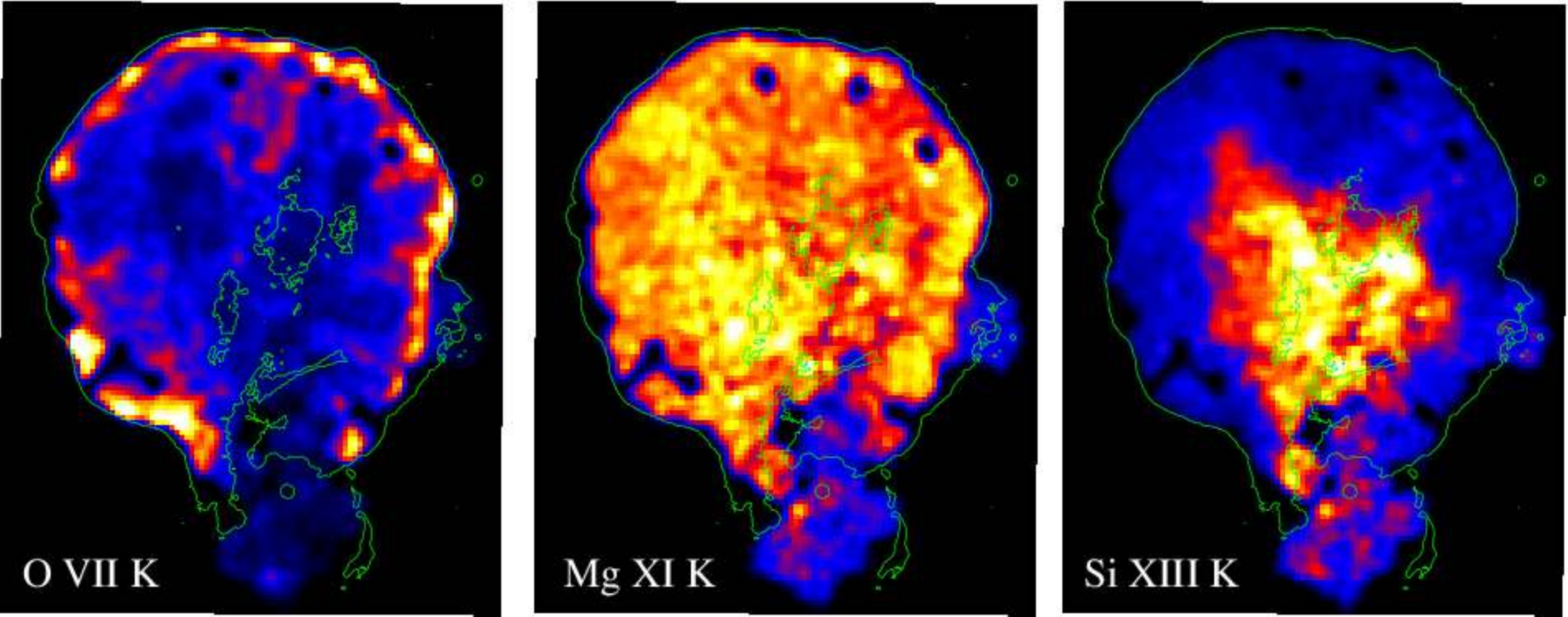}
\caption{\small{Equivalent width maps for K$\alpha$ lines of O VII, Mg XI, and Si XIII \citep{uchida09}.   There are still some regions left unobserved with {\it Suzaku}, which appear as dark spots.}} 
\label{EMimage} 
\end{figure}

The first high resolution X-ray imagery of the Cygnus Loop was obtained with 
{\it Einstein} \citep{ku84}, and nearly complete maps of the remnant were made 
with {\it ROSAT} \citep{aschenbach99}, {\it ASCA} \citep{miyata98}, 
{\it XMM-Newton} \citep{tsunemi07} and {\it Suzaku} \citep[][and references therein]{uchida11}.  
Figure~\ref{fig:XAcyg_image} (a) shows the X-ray image of the Cygnus Loop obtained with {\it ROSAT}.    The the {\it ROSAT} HRI image mainly shows the low $kT_ e$ component of X-ray emission.   
The spectrum is dominated by emission lines from various elements: O, Ne, Mg, Si, S, and Ar.   
The improved spectral resolution of X-ray CCDs made possible the production of images in some of these emission lines.  Some {\it Suzaku} line images are shown in Figure~\ref{EMimage}.
The O line dominates in the shell region indicating an ISM origin of the shocked gas there, while Si fills the interior indicating the hot ejecta component.    

\subsubsection{Cygnus Loop - Solving the low abundance problem}

The basic question is what causes the ``low abundance'' problem.  If abundances in the post-shock plasma are actually low, then accurate abundance measurements should provide an important clue as to the reason.  On the other hand, if the problem arises due to physical processes that are not included in the standard analysis (of CCD data), then this should also be apparent.

\begin{figure}[htb]
\begin{center}

\includegraphics[scale=0.57,angle=0]{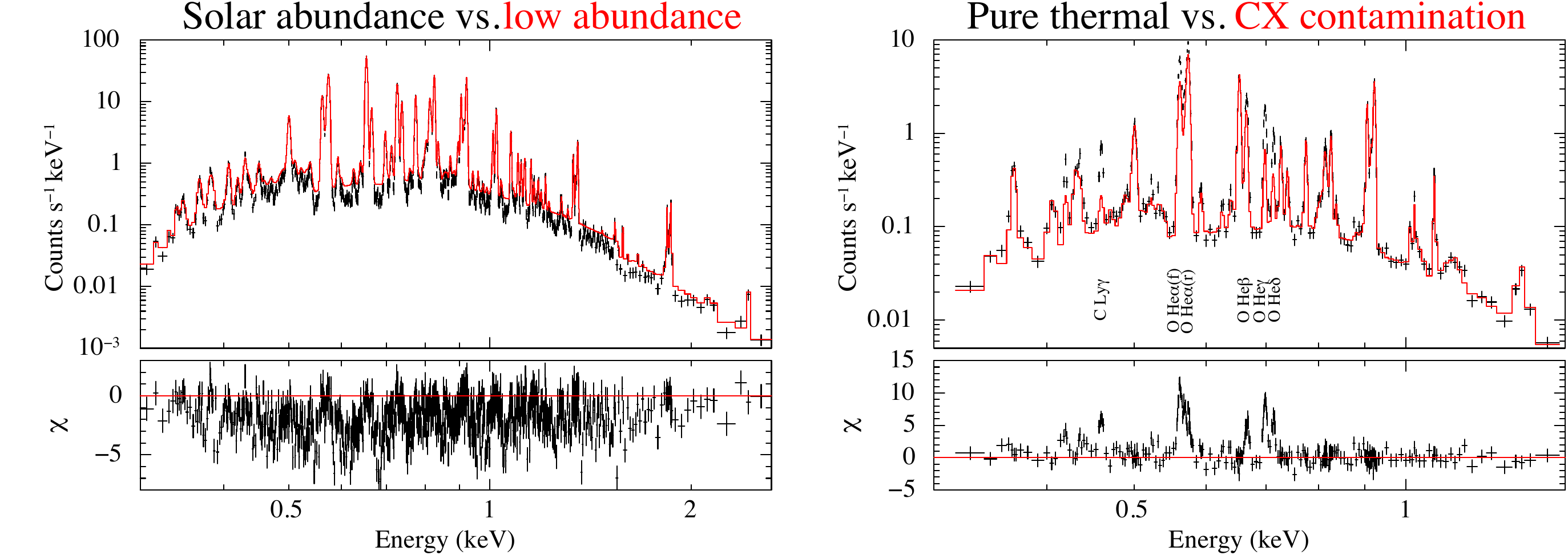}
\caption{Left: Simulated 50~ks SXS spectrum for the northeastern region of the 
Cygnus Loop (the region is indicated by arrow 1 in 
Figure~\ref{fig:XAcyg_image}).  
Red and black correspond to the low abundance case ($\sim0.2$\,solar) 
and the solar abundance case, respectively.  The lower panel shows the 
residuals for the simulated data and the model for the low-abundance case.
Right: Simulated 100~ks SXS spectrum for the northeastern edge of the Cygnus 
Loop (the region is indicated by arrow 2 in Figure~\ref{fig:XAcyg_image}).    
The simulated SXS spectrum includes contamination by CX emission, 
while the model represents pure thermal emission. \label{fig:NEcyg_spec}}
\end{center}
\end{figure}

We first demonstrate that it is possible to accurately characterize the plasma with {\it ASTRO-H}.  This is an important first step because by examining line strengths of individual lines we can verify whether the models applied to the CCD data are appropriate.   For this, we  simulate a 50 ksec observation of the bright 
northeast rim of the Cygnus Loop, as shown in Figure~\ref{fig:XAcyg_image}.  
To reduce the effects of CX emission, we select a region 
 $\sim6^\prime$ inside from the outer rim.  The left panel of Figure ~\ref{fig:NEcyg_spec} shows a comparison between models with with low abundances 
($\sim0.2$\,solar) and solar abundances.  Analysis of the simulated spectra
shows that we can measure absolute abundances of C, N, O, Ne, and Mg individually to 10\% 
accuracy (90\% confidence level), something that is impossible with CCD spectra.

\begin{figure}[htb]
\begin{center}
\includegraphics[width=0.5\textwidth]{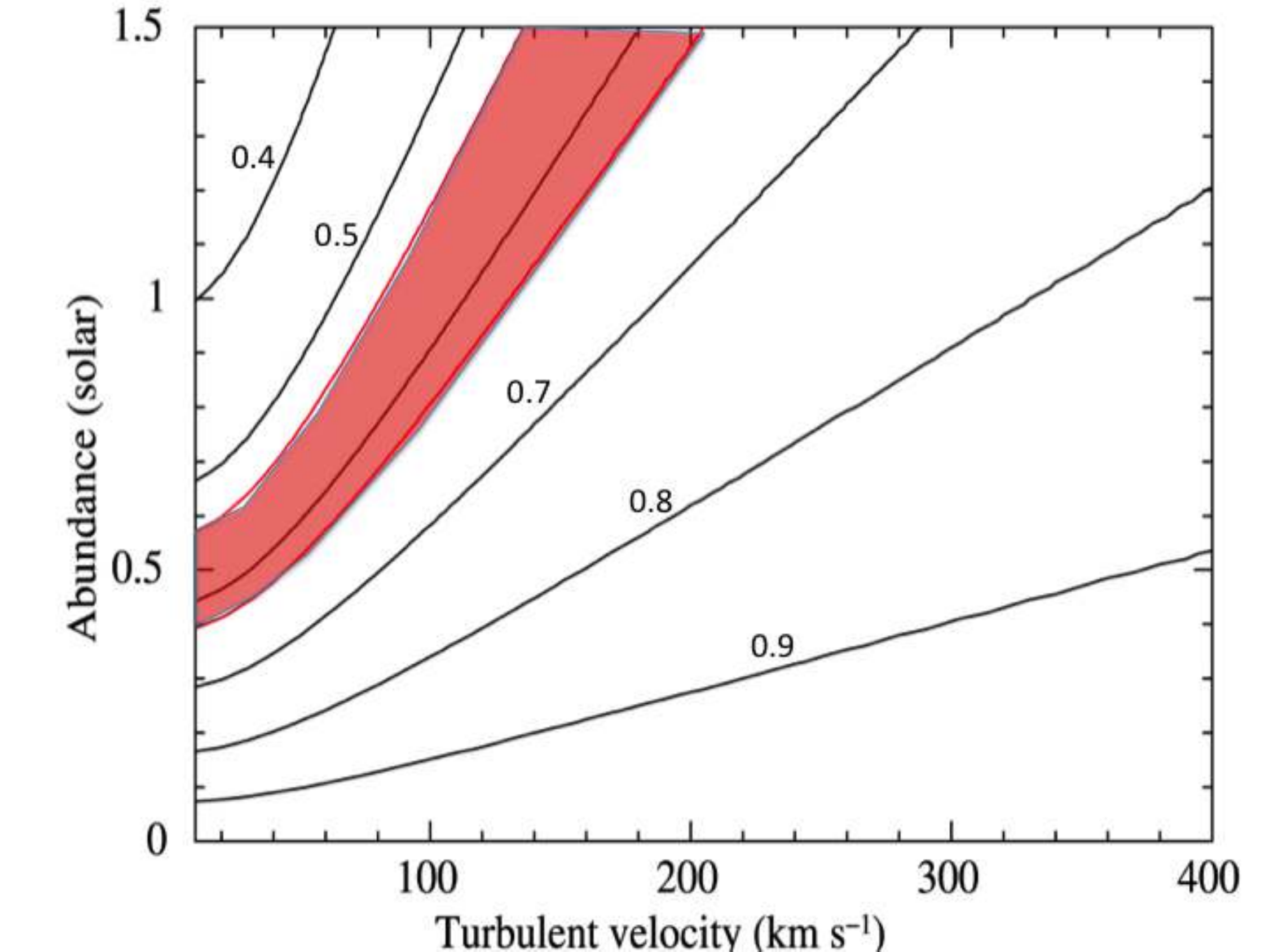}
\caption{The escape probability of the resonance line in O He$\alpha$ (Z-axis) as a function of the O abundance (Y-axis) and the turbulent velocity (X-axis).  The red area indicates a 90 percent confidence limit (statistical uncertainty) region for a 20 ks exposure time at the 0.95 R$_{sh}$ position.}
\label{fig:cygloop_turb_mod}
\end{center}
\end{figure}

Secondly, to answer the question of whether atomic physics is the root cause of the apparently low abundances observed in the Cygnus Loop and other older SNRs, we need to obtain high-S/N and high-resolution observations of several positions along the rim of the Cygnus Loop.  These observations are needed to fully characterize the spectrum, and to separate questions of abundances from those associated with other physical processes such as charge exchange and resonance scattering.  Relatively short exposures in various regions of the Cygnus Loop with {\it ASTRO-H} can be used to evaluate the importance of these processes.   

Resonance scattering changes the ratio of forbidden and resonance lines, and by reducing the intensity of the resonance lines can result in apparently lower abundances than is actually the case.  And example of this is shown in  Figure \ref{fig:cygloop_turb_mod}, which shows the escape probability for the O He-like resonance line at 0.57 keV in a plane of the O abundance and the turbulent velocity.  

To demonstrate the ability of {\it ASTRO-H} to measure the effects of atomic processes that can affect abundance measurements, we have simulated short observations at two positions in the Cygnus Loop, one immediately behind the shock and one slightly further in.  Both positions are simulated assuming $kT_e$ of 0.25 keV, an ionization age of 1 $\times$10$^{11}$~s~cm$^{-3}$, and solar abundances.  However, the line of sight column densities differ:  0.1$\times$10$^{20}$~cm$^{-2}$ for the field at the edge, 0.25$\times$10$^{20}$~cm$^{-2}$ for the interior field.

\begin{figure}[htb]
\begin{center}
\includegraphics[width=16cm]{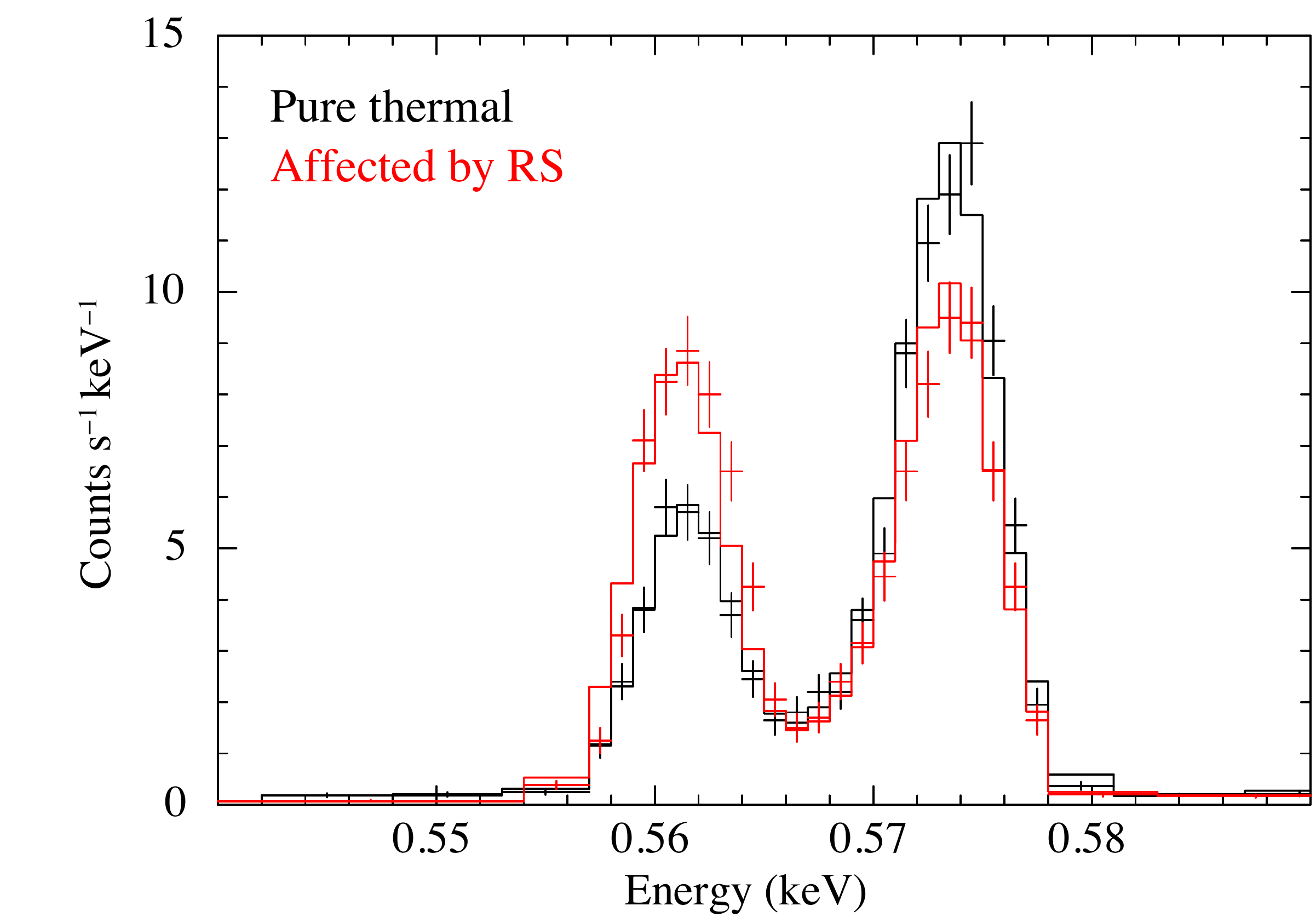}
\caption{A  30 ks simulation of the NE limb of the Cygnus loop, somewhat inside the shock (at $R/R_{\rm s}$ of 95\%), showing the effects of resonance scattering on the spectrum.  The black data points are for a model in which charge resonance scattering is ignored; the red points include the effects of resonance scattering. }
\label{fig:cygloop_fr}
\end{center}
\end{figure}

Our simulations show that we can accurately measure forbidden to resonance line ratios in both cases, to an accuracy of better than 10\%, which will allow a direct assessment of the importance of resonance scattering in estimates of abundances in old SNRs like the Cygnus Loop.  We also can distinguish between the case where charge exchange is important and not.

\subsubsection{Cygnus Loop -- Shock Cloud Interactions}

The Cygnus Loop, Puppis~A, and the Vela SNR provide nice 
examples of cloud-shock interactions.  We here examine the so-called ``XA knot" along the eastern 
edge of the Cygnus Loop, (see, 
Figure~\ref{fig:XAcyg_image}), where detailed optical and 
X-ray studies have been already performed 
\citep[e.g.,][]{Danforth2001,Miyata2001,Zhou2010,McEntaffer2011}. 
As shown in Figure~\ref{fig:XAcyg_image} (c), the knot shows a chaotic network of 
optical filaments, providing both edge-on and face-on views of shock 
fronts propagating into dense clouds.  The speeds of the shocks
have been estimated to be $\sim$100--200\,km\,s$^{-1}$.  Given this
complex morphology, an SXS measurement of this region will sample a variety of shock speeds, yielding a spectrum suggesting random gas (bulk) motions of $\sim$400\,km\,s$^{-1}$,
i.e., $\pm$200\,km\,s$^{-1}$.  Moreover, plasma instabilities along the interface between the clouds and intercloud medium would 
yield additional turbulent velocities of $\sim$10\% of the shock speed 
\citep{Pittard2010}.  Given these considerations, we would expect a 
random gas motion of at least $\sim$440\,km\,s$^{-1}$, which causes 
line broadening of $\sigma\sim$1\,eV at O VIII Ly$\alpha$ (654\,eV).  
We introduce this amount of broadening in the following SXS simulation 
for cloud-shock regions.

\begin{figure}[htb]
\begin{center}

\includegraphics[scale=0.6,angle=0]{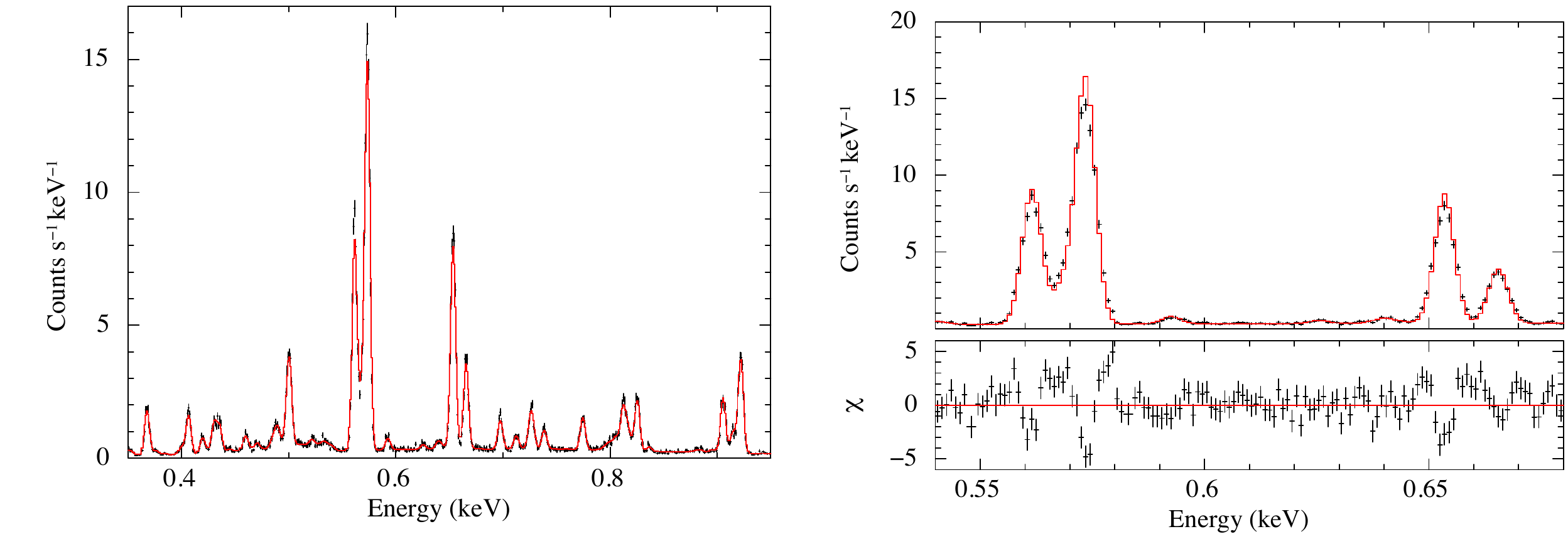}
\caption{Left: Simulated 20\,ks SXS spectrum for the XA region of the 
Cygnus Loop.  
Right: Close-up of the SXS spectrum in the left panel, 
but with the best-fit model having no line broadening, i.e., 
only instrumental broadening.  The lower panel shows residuals.
It is clear that the line profiles in data are broader than those in the model. \label{fig:XAcyg_spec}}
\end{center}
\end{figure}

Here we  demonstrate that {\it ASTRO-H} can be used to detect broadening in the lines. 
For this, we select the brightest area in the XA region, 
as shown in Figure~\ref{fig:XAcyg_image} (b).  To simulate SXS spectra, 
we first extracted an {\it XMM-Newton} MOS spectrum, and then fit it with a 
non-equilibrium ionization (NEI) plasma model.  Based on the best-fit emission 
model, for which line broadening of 440\,km\,s$^{-1}$ (or $\sim$1\,eV at 
654\,eV) estimated above is taken into account, we simulate an SXS 
spectrum as shown in Figure~\ref{fig:XAcyg_spec}.   Due to  the
brightness of this region, an exposure time of only 20\,ks allows us to collect 
$\sim$13,000\,counts in 0.2--2\,keV.  In the right panel of 
Figure~\ref{fig:XAcyg_spec}, we show a close-up spectrum for O K-shell 
lines, in which we show an emission model (in red) having no line
broadening (i.e., the broadening is solely due to instrumental 
resolution).  It is clear that the simulated spectrum has broader lines 
than the model, and thus we can measure the line widths (and turbulent 
velocities in the region) within about 10\% accuracy.  It should be noted that detecting such minor
broadening will require that the SXS spectral resolution is very well calibrated.

\subsubsection{Cygnus Loop -- Nature of the progenitor}

\cite{preite11} argue, essentially on energetic grounds, that the Cygnus Loop was the result of 
explosion of a relatively low (8--10~\MSOL{}) mass star, which produced a SN of low energy,  
possibly an electron-capture SN.  On the other hand, \cite{uchida11} argue from {\it Suzaku} spectra 
that abundances of Ne, Mg, and Si relative to O are better understood in terms of 
a 12--15~\MSOL{} progenitor.   Using {\it ASTRO-H}, it will be possible to accurately measure 
the metal abundances for direct comparison with models such as those of \cite{woosley95}.

\begin{figure}[h]
\includegraphics[width=16cm]{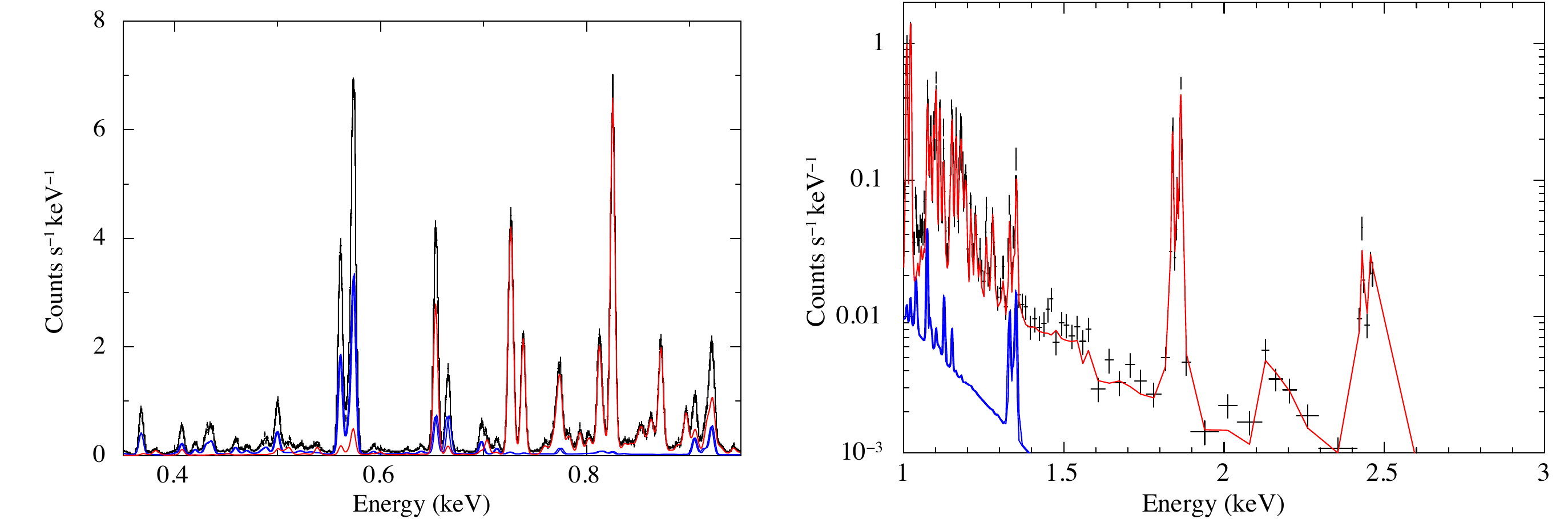}

\caption{\small{Simulation of a 100\,ks SXS observation of the interior of the Cygnus Loop.  Left shows the spectrum below 1\,keV where blue and magenta show two shell component: near side and far side.  
Red shows the ejecta component.  
Right shows the spectrum above 1\,keV where ejecta the component dominates.}}
\label{CygSpectrum} 
\end{figure}

Figure~\ref{CygSpectrum}-left shows a 100\,ks  SXS simulation of the center of  the Cygnus Loop.  The emission here consists of three components.  Blue and magenta show two shell components, near side and far side, respectively.  Red shows the ejecta component while the black shows the sum.  
As shown in Figure~\ref{CygSpectrum}-right, the spectrum above 1\,keV mainly comes from the ejecta component.  

Although we cannot separate the motion between the near side shell and the far side shell, the O VIII line width of might provide a measure of the expansion velocity.  The Si XIII triplet will be resolved, allowing us to determine the plasma conditions.  The simulation shows a representative interior pointing.  

\begin{figure}[htb]
\begin{center}
\includegraphics[width=0.49\textwidth]{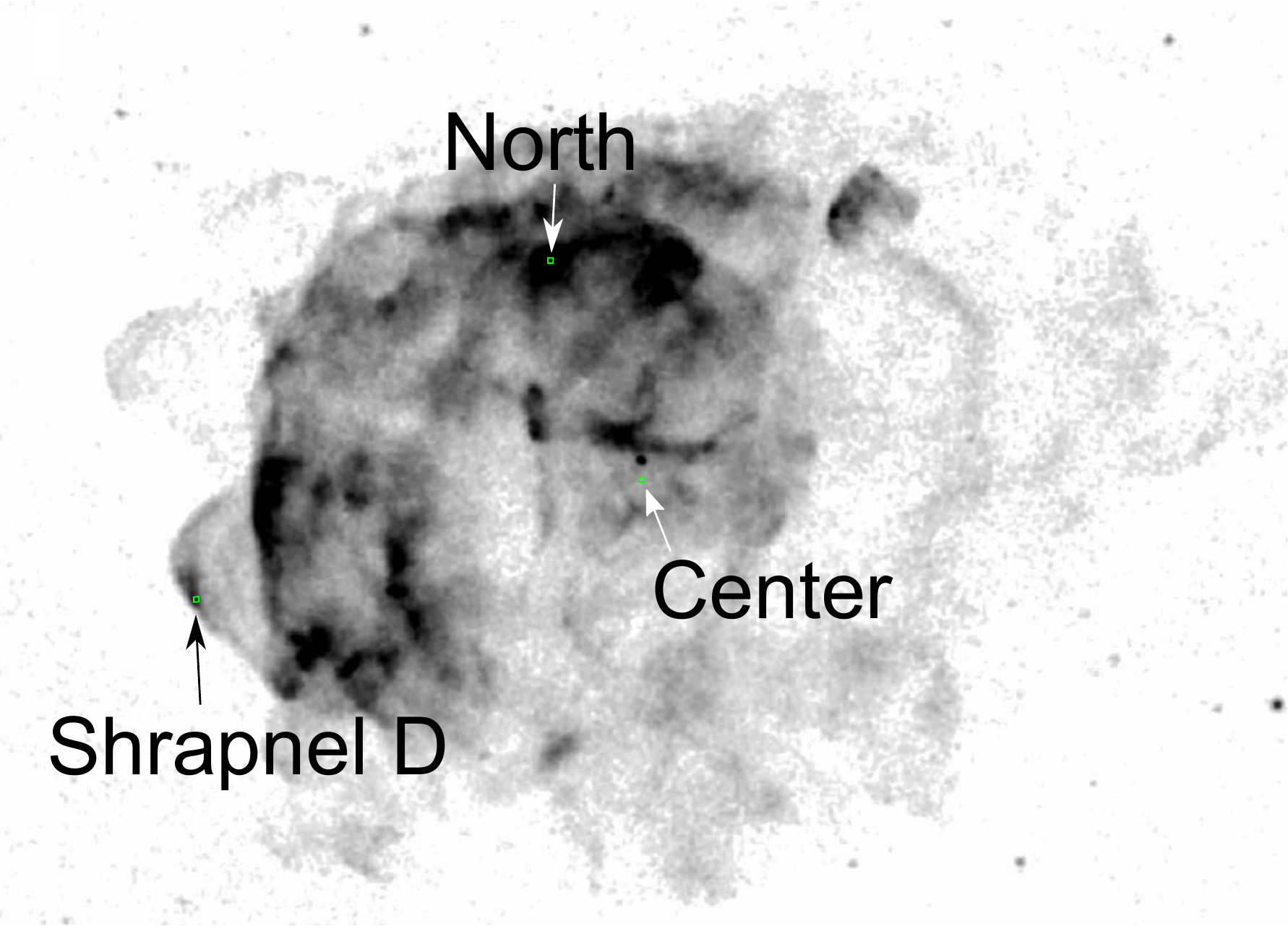}
\includegraphics[width=0.49\textwidth]{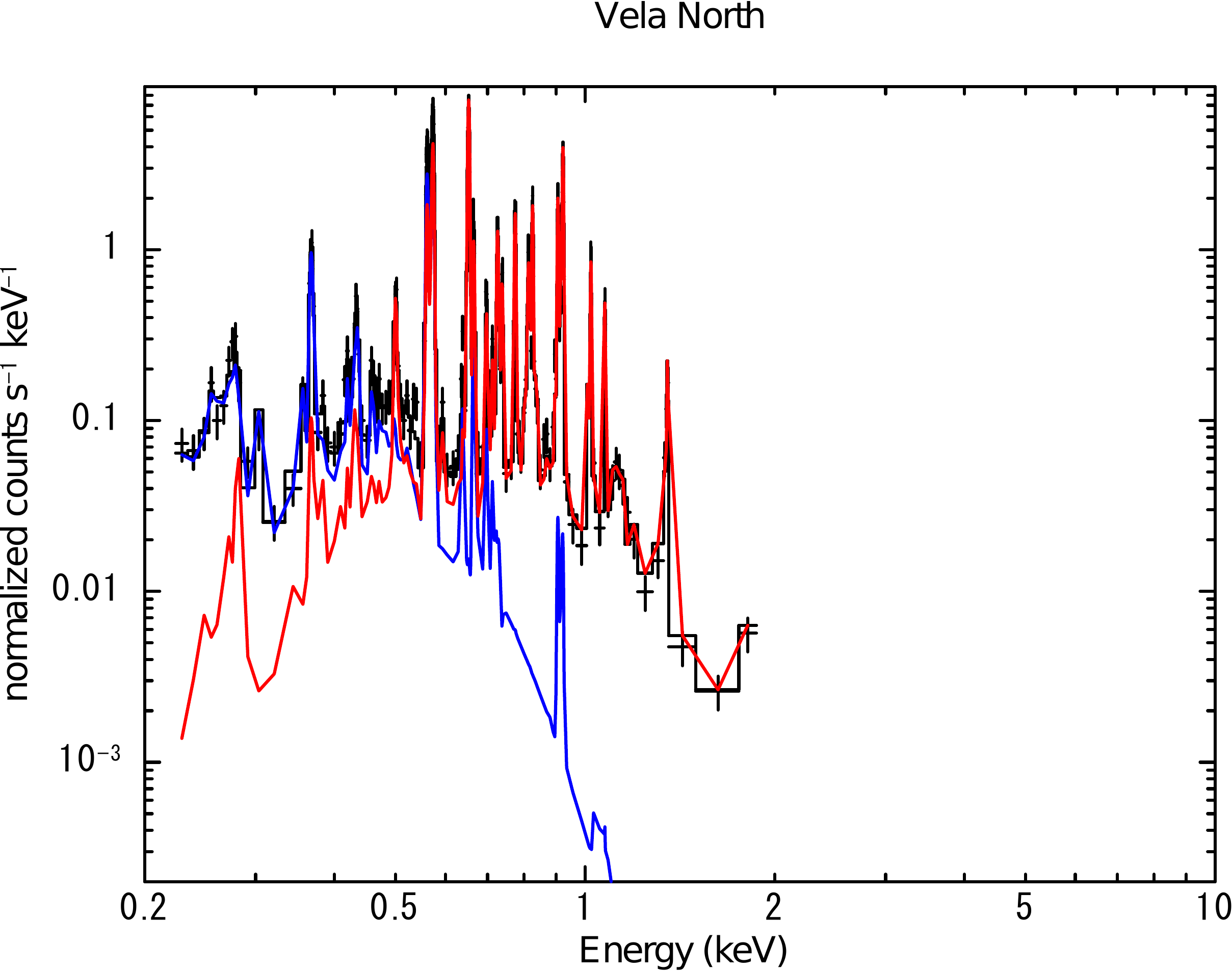}
\includegraphics[width=0.49\textwidth]{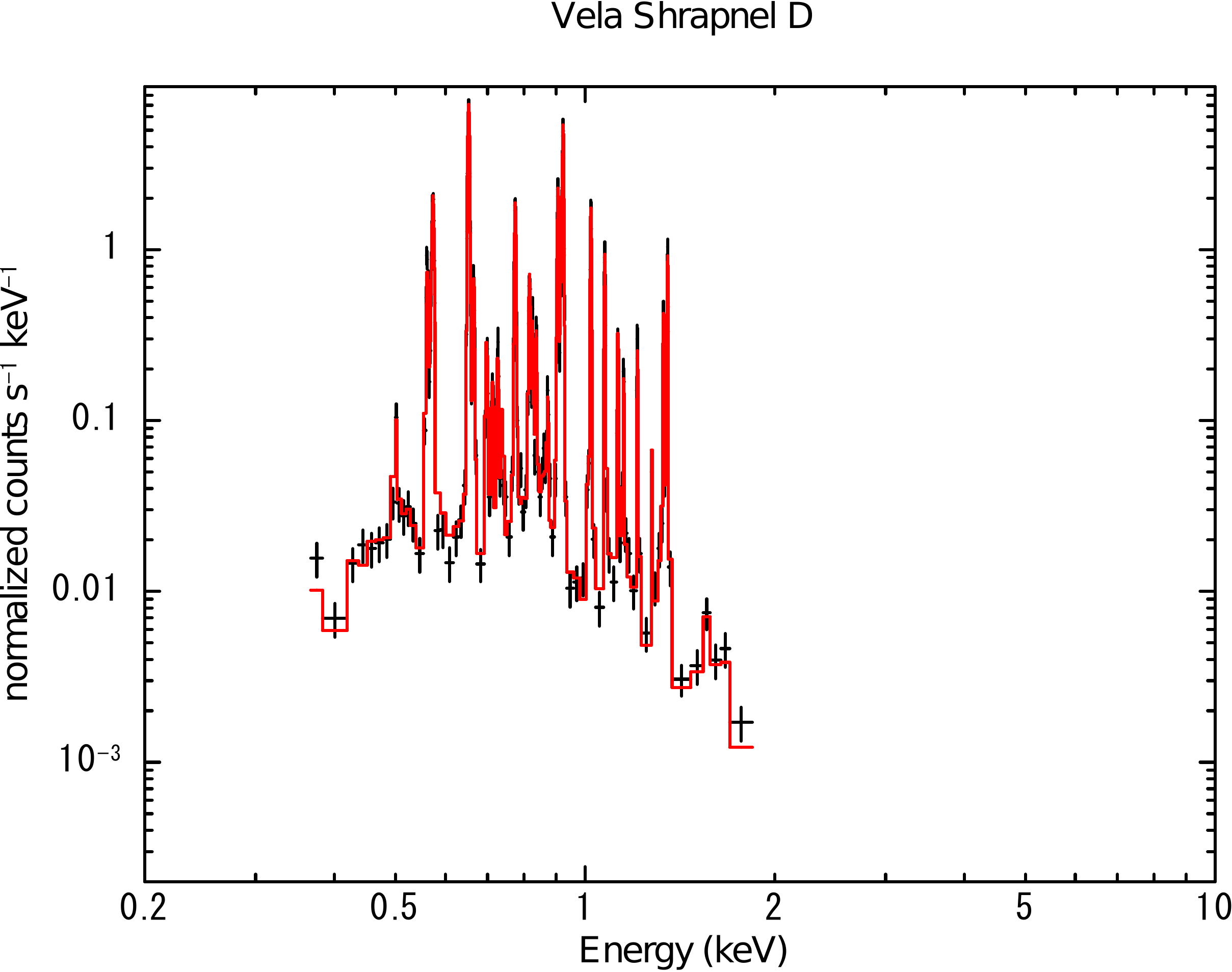}
\includegraphics[width=0.49\textwidth]{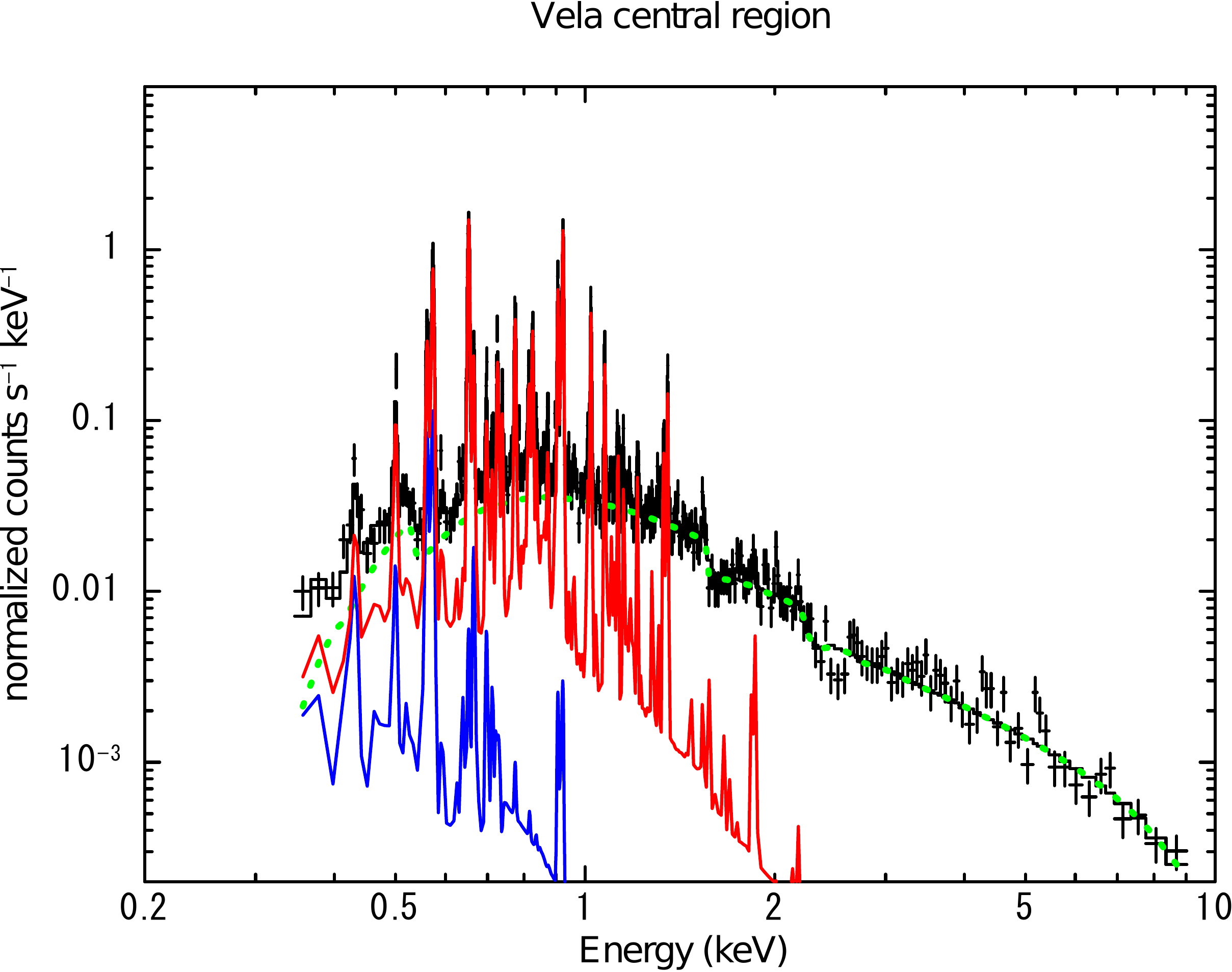} 

\caption{{\bf Top-left}: The ROSAT All-Sky Survey image of the Vela SNR. The north,
 shrapnel D, and center regions, for which we perform SXS simulations, are
 shown with green boxes. {\bf Top-right, bottom-left, and bottom-right}: The SXS
 simulations of the north, shrapnel D, and center regions. Red, blue, and green
 lines represent the high-, low-temperature, and non-thermal plasma models,
 respectively. }

\label{fig:vela}
\end{center}
\end{figure}

\begin{figure}[htb]
\begin{center}
\includegraphics[width=10cm]{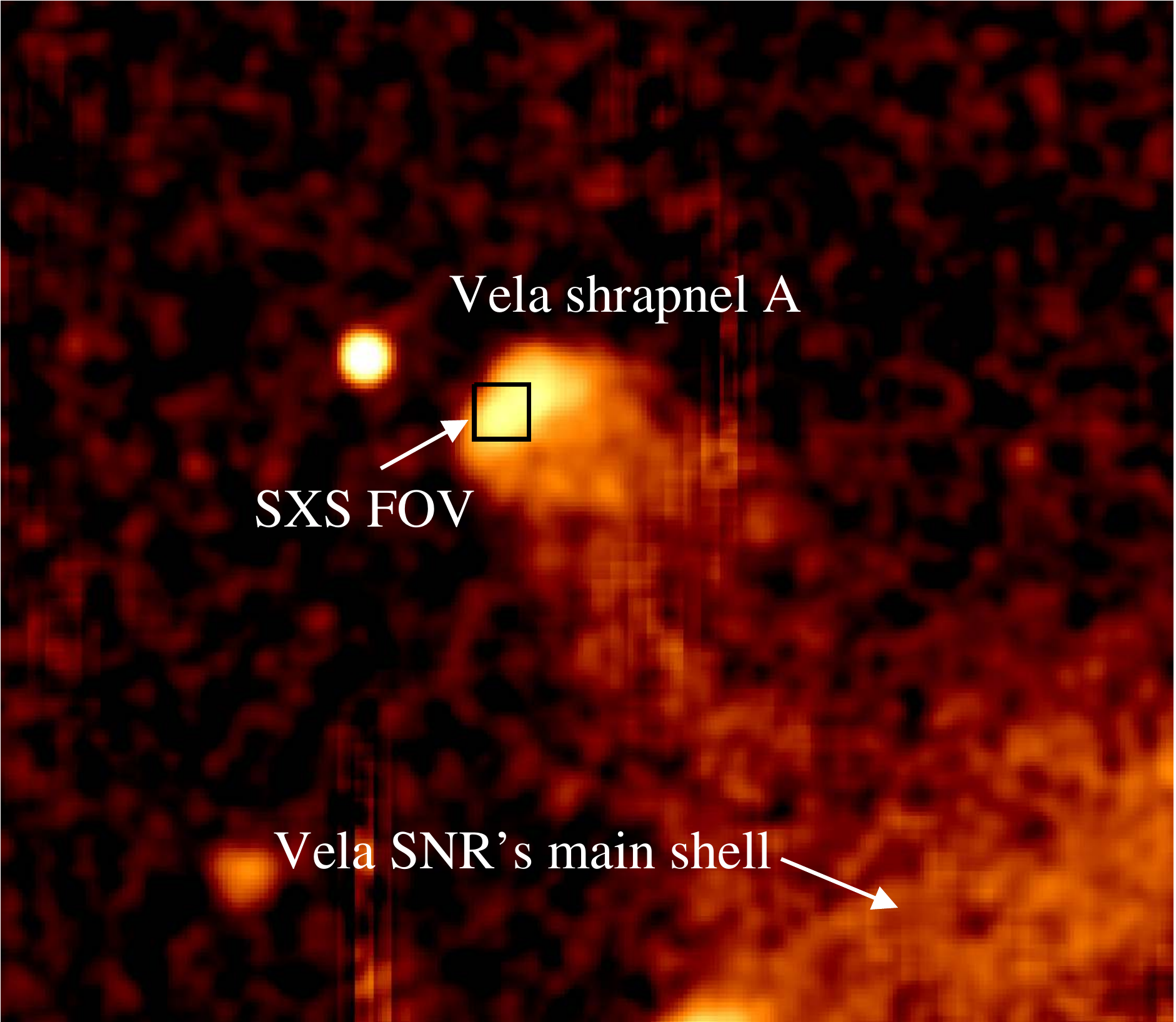}
\caption{
ROSAT All-Sky Survey image of Vela shrapnel A. The SXS FOV is
shown as a black box.
\label{fig:vela_detail}
}
\end{center}
\end{figure}

\begin{figure}[htb]
\begin{center}
\includegraphics[width=16cm]{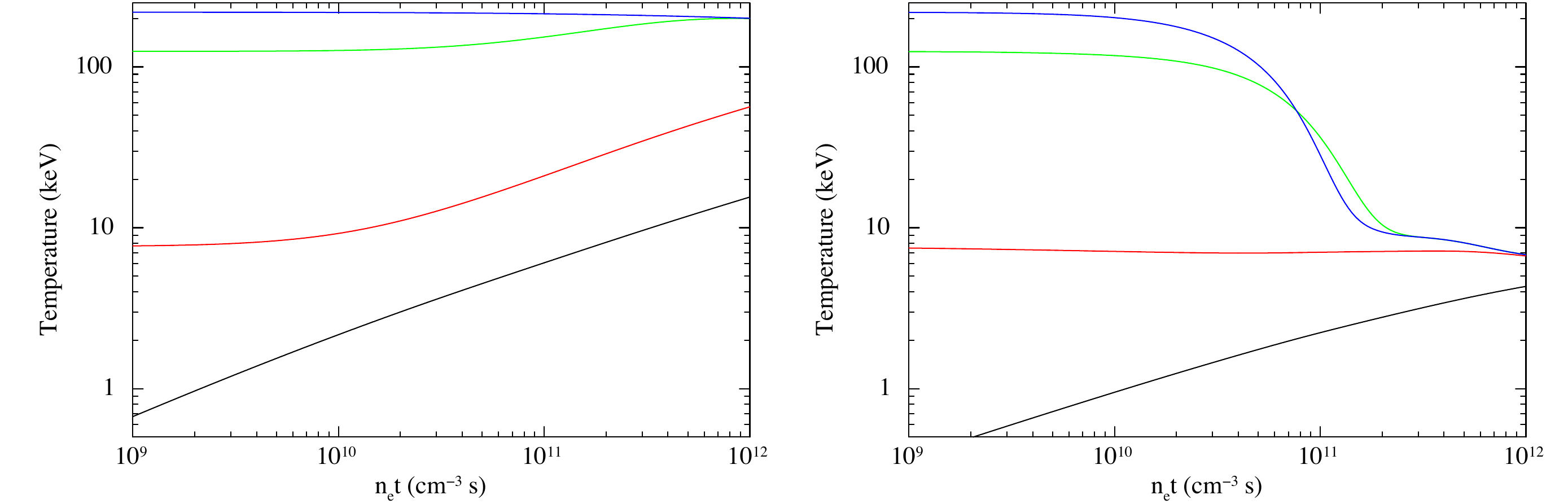}
\caption{
left: Temperature histories for electrons (black), protons (red), oxygen
(green), and silicon (blue) as a function of the ionization timescale.
In the calculation, we assume that the plasma is (nearly) pure metal
abundances and that temperature equilibration takes place through only
Coulomb interactions after collisionless shock heating
($V_{shock} \sim 2000$~km~s$^{-1}$).
right: Same as left but for a slightly elevated abundance (several-solar
abundance) plasma.
}\label{fig:v500}
\end{center}
\end{figure}

\subsubsection{Vela SNR - Background }

Another old ISM-dominated SNR that is important to study is the Vela SNR.  It is a textbook example of an evolved core-collapse SNR,
exhibiting overwhelming soft X-ray emission from blast-wave shocked clouds
\citep[e.g.,][]{lu00}, underlying relatively-faint line emission from metal-rich
ejecta \citep[e.g.,][]{lamassa08}, and interaction between an expanding pulsar
wind nebula and a confining reverse shock \citep[e.g.,][]{blondin01}. In addition,
there are a few isolated structures called ``shrapnel'' beyond the forward shock, which are identified as ejecta fragments by their strong line
emission by metals thought to be ejecta\citep[e.g.,][]{katsuda05}. 

Figure~\ref{fig:vela} shows SXS simulations of three representative regions: north
(30 ks), center (100~ks), and shrapnel D (80~ks), which are intended to indicate the breadth of studies that can be carried out in the Vela SNR. The exposure times are chosen so
that statistics for the spectra are roughly the same. The thermal spectra within the main body of
the Vela SNR are generally well described by a two-temperature model
\citep{lu00}. The low-temperature plasma (blue in the simulation spectra) arises in the 
shock-heated ISM clouds that dominate the ROSAT image shown in 
top-left panel of Figure~\ref{fig:vela}, while the high-temperature plasma (red in the
simulation spectra) originates from either the evaporation of shocked clouds,
ejecta or a mixture of the two \citep{miceli06}.  {\it ASTRO-H} spectra are needed to advance our understanding of the physical processes and abundances in each of these regions.

The north region is dominated by the soft X-ray emission arising
from the interaction between the blast wave and ISM clouds
\citep[e.g.,][]{miceli05}.   Its high surface brightness will allow the obtaining of high S/N spectra with relatively short expoosures.  Although extensive CCD-based studies have explored cloud--shock interactions within this region \citep{lu00, miceli05, miceli06}, its low temperature
($\sim$0.1~keV) has hampered detailed plasma diagnostics to date,  because the emission
lines, mostly below 0.5~keV. The {\it ASTRO-H} SXS will isolates C, N, and O lines for the first time and
thus provide new insight into the dynamics and plasma conditions in the shock-heated
clouds.

The central region of the Vela SNR contains not only soft
X-ray emission from shocked clouds, but also line emission from metal-rich ejecta and non-thermal
emission from the pulsar wind nebula
\citep{lamassa08, katsuda11}. Since the ejecta at the center of the remnant is
expected to be less contaminated by ISM in comparison with those at the shell, observations of this plasma will allow {\it ASTRO-H} to study the abundance pattern in the ejecta, which is needed for better understanding of
the progenitor. However, in most locations, other thermal and non-thermal components
coexisting in projection partially hide or dilute the ejecta emission.  As result, observations of the almost purely ejecta-dominated shrapnel D region are critical.   The SXS will determine with unprecedented precision the plasma
conditions in both shrapnel D and the high-temperature component in the central
region, thereby allowing efficient extraction of abundance information.
These measurements will allow us to determine whether the high velocity ejecta
overrunning the blast wave and the slow velocity ejecta in the interior 
originate from the same progenitor layer, thus shedding light on the 
explosion mechanism.

Given Vela's age (over 10,000 yr), an interaction between the pulsar wind
nebula and the reverse-shock is expected to have occurred \citep{blondin01}.  The filamentary structure of the radio morphology is highly suggestive of mixture of two different fluids,
namely pulsar wind and ejecta \citep{bock98}. The situation where the strong
magnetic fields, $\sim$100~$\mu$G at least around the bright PWN \citep{pavlov03},
and the thermal plasma interact with each other may produce a supra-thermal
electron distribution, deviating from Maxwellian  \citep{masai02, kaastra09}. 
The supra-thermal electrons will change the
relative intensities of the satellite lines \citep{kaastra09}, which may be tested
with the SXS spectrum.

\begin{figure}[htb]
\begin{center}
\includegraphics[width=16cm]{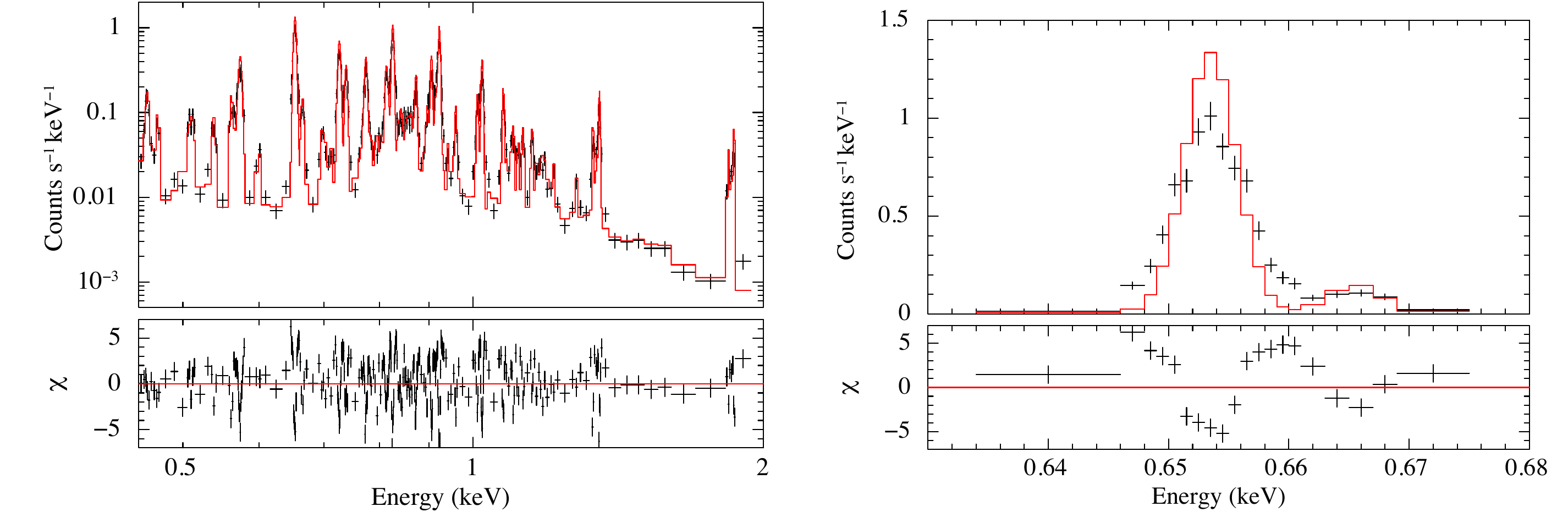}
\caption{Left: Simulated 200~ks of Vela A with SXS. The simulated data in black
assume a pure metal case, while the model curve in red assume several
solar abundances. Right: Close-up of the O~Lyman band.
We can measure the line width to be 2.4$\pm$0.2~eV.
}
\label{fig:vela_d_models}
\end{center}
\end{figure}

\subsubsection{Using velocity-broadening to measure ejecta metallicity in the Vela SNR through observations of ``shrapnel''}

A number of older SNRs, including Vela,  show evidence of metal-rich plasmas, and
are worthy of study with {\it ASTRO-H}. To date, we have only lower limits
on the abundances for such high-metallicity plasmas.  As we have discussed above, with the ASTRO SXS we will be able to carry out straightforward abundance studies of many of these SNRs, and this will be very important to determining the nature of the progenitor of these objects.  However, {\it ASTRO-H} SXS also 
enables a different kind of test of the amount of overabundance by using the 
 the width of the lines to estimate the post-shock ion temperature of the gas after Coulomb equilibration, which depends on a combination of the shock speed and the metallicity. 
 
 The Vela SNR's Shrapnel A, which is a
relatively compact, Si-rich ejecta knot (see, Figure 9), provides the
best place to pursue such an observation.
The basic idea is as follows: For a fixed shock speed, all elements have
the same initial shock-heated temperatures which are proportional to
their elemental masses (in collisionless shocks). However, the temperature
evolution due to Coulomb equilibration depends on the metallicity.  For fixed shock speed, the temperature of the plasma 
is much higher in the case with pure metals
than one with a mixture of H and heavier metals.  For example, as shown in Figure 10,
for an assumed shock velocity of 2000 km\,s$^{-1}$ (which is a mean
velocity, radius/age, for Vela Shrapnel A), the ion temperature is expected
to be about 200\,keV for a pure metal case, but only 7\,keV for a
several-solar case. The temperature of either 200\,keV or 7\,keV
would result in line broadening of either $\sigma = 2.4$\,eV or
0.44\,eV, respectively, at the ionization timescale in Vela A
\citep[($n_{\mathrm e}t \gtrsim
10^{11}$\,cm$^{-3}$\,s:][]{Tsunemi1999,Miyata2001,Katsuda2006}.
Such a difference can be easily distinguished by the SXS with a 200 ks
exposure, for which the SXS simulation is presented in Figure  \ref{fig:v500} , hence
we will be able to determine the absolute metallicity there.

\section{Mixed-Morphology Supernova Remnants \label{mixed_morph}}

\subsection{Background and Previous Studies}

There exists a substantial population of old SNRs displaying centrally peaked X-ray emission arising from thermal plasma \citep{white91, rho98}.  These remnants generally appear shell-like in the radio, hence the name mixed-morphology SNRs \citep{rho98}. They are generally found near dense interstellar structures, usually molecular clouds, often interacting with them.  A large fraction contain OH masers \citep{yusef03}.  

Recent studies of mixed-morphology SNRs have revealed three important new features.  
At least one-third of these SNRs have associated $\gamma$-ray emission \cite[see, e. g.] []{vink12}. 
This emission is thought to arise from relativistic protons, originally accelerated in 
the magnetic field of the forward shock, colliding with material in the nearby 
molecular clouds.  The resulting $\gamma$-ray emission is the strongest evidence
favoring cosmic ray acceleration in SNR shocks.  A second recently discovered feature 
is the presence of enhanced metal abundances in the interiors of about half the known 
mixed-morphology SNRs \citep{lazendic06,shelton04}.  These enhanced abundances 
suggest that ejecta has not completely mixed with the surrounding ISM, despite the fact 
that most are of these SNRs are thought to be relatively old ($>$20,000 years), and that 
many are in the radiative phase.  Third, while early studies of these remnants indicated 
their X-ray emitting plasma is close to collisional ionization equilibrium \citep{rho98}, 
observations using {\it ASCA} and {\it Suzaku} have revealed surprisingly that some contain over-
ionized and thus recombination-dominated plasma (RP), evidenced by both line ratios 
\citep{kawasaki02} and radiative recombination continua (RRCs: \citealt{yamaguchi09}).  

The mechanisms responsible for forming mixed-morphology SNRs are still not completely 
understood.  The absence of an X-ray shell suggests that these remnants are dynamically 
old, and that the forward shock has slowed down to a velocity too slow to produce 
substantial X-ray emission above $\sim$1 keV.  However, most of these remnants are 
observed through high column density (1--2$\times$10$^{22}$ cm$^{-2}$), so X-ray 
emission from a shell produced by a forward shock with velocity less than a few hundred 
km s$^{-1}$ would be absorbed.  However, the only known Galactic mixed-morphology 
remnant with a low intervening column, G65.3$+$5.7, shows a shell only below $\sim$0.5 
keV, supporting the idea that mixed-morphology remnants have evolved to an 
evolutionary phase later than the old shell-like SNRs typified by the Cygnus Loop.  
Indeed, some are thought to have entered the radiative phase \citep{shelton04}.  

\cite{white91} suggested that the centrally enhanced emission arose from clouds 
bypassed by the primary shock due to their high density, which subsequently were 
evaporated by hot gas in the interior.  Their simple self-similar model for the evolution of 
such SNRs, which predicts younger ages and lower swept up masses than a Sedov-
based single component ISM analysis, is almost certainly incorrect in detail, but whether 
evaporating clouds contribute to the interior gas density is less clear.  \cite{shelton99} 
suggested instead that a SNR expanding into an dense ISM with a strong decreasing 
density gradient, coupled with thermal conduction or turbulent mixing, was more likely to 
produce the observed morphology.  Neither model includes the possibility that emission 
from ejecta could contribute to the X-ray morphology of these SNRs.  

Recent {\it Suzaku} study of mixed-morphology SNRs 
\citep{yamaguchi12, sawada12, uchida12} suggests that a specific sequence of events 
could be responsible for producing the RP.  The shock from a supernova exploding in a 
dense circumstellar medium (CSM), presumably formed by a pre-supernova wind, would 
rapidly heat both the CSM and ejecta to a high temperature ($kT_{\rm high}$).  The high 
density causes the plasma to rapidly approach collisional ionization equilibrium.  As the 
forward shock expands beyond the CSM envelope, the shock rapidly accelerates. The 
shocked material becomes rarified and cooled ($kT_{\rm low}$), and if the adiabatic 
cooling timescale of the electrons is shorter than the recombination timescale of the ions, 
then an over-ionized plasma is formed. It can persist for a long time, while gradually 
relaxing (recombining) to reach collisional ionisation equilibrium at $kT_{\rm low}$. 

This model, originally proposed by \cite{itoh89}, raises as many questions as it purports to 
solve, many of which are testable.  It requires a specific pre-SN wind structure, and thus 
only progenitors in a specific mass range would produce mixed-morphology remnants.  It 
is not clear how the ejecta remain centrally concentrated, although some idea has been 
proposed to explain the morphology as well \citep{shimizu12}.  Perhaps more importantly, 
all mixed-morphology remnants are dynamically old; where are the early stage mixed-
morphology remnants and where are the dynamically old remnants from outside this 
mass range?  Given the morphological complexity of these remnants and their 
environments, it is likely that no single model works, and rarefaction, conduction, and 
evaporation (and possibly other mechanisms) all contribute.

\subsection{Prospects \& Strategy}
There are two primary objectives in studying mixed-morphology SNRs 
using {\it ASTRO-H}.  The first is a thorough study of the over-ionized plasmas.  The  SXS 
will provide powerful plasma diagnostics, allowing us to carefully measure line ratios from 
a variety of metals, and therefore better quantify the ionization conditions.  For remnants 
in which strong RRCs have been detected, it can be used to precisely measure known 
RRCs and sensitively search for RRCs from different metals.  SXS measurements will 
allow us to determine whether the over-ionization conditions vary across a remnant.  
Additionally they will enable a more sensitive search for RRCs and line ratios indicating 
over-ionization in remnants where these features have not been detected.  

The second objective is to more accurately characterize the abundances and distribution 
of the ejecta observed towards the centers of many mixed-morphology remnants, with the 
goal of establishing progenitor masses.  The relationship between the presence of over-
ionization and enhanced abundances will provide clues about the progenitor and the 
conditions responsible for forming mixed-morphology SNRs.

Both objectives have the potential to provide new insights into the origin and evolution of 
this class of SNRs.  A carefully constructed observing program can fulfill both objectives 
with a common set of observations, especially as several mixed-morphology SNRs show 
both traits.  Most of these remnants are much larger than the SXS field of view and have 
relatively low surface brightness.  Complete mapping in general would not represent a 
productive use of {\it ASTRO-H} time.  However, all have been mapped using previous 
missions, and these maps can be used to define a small set of pointings for each remnant 
of interest.

If the rarification scenario is correct, then understanding the early evolution of the plasma 
is critical as it would naturally be related to the explosion mechanism and environment of 
mixed-morphology SNRs.  The SXS enables independent determination of key 
parameters associated with the initial explosion environment and subsequent evolution, 
namely the electron temperatures before and after the cooling ($kT_{\rm high}$ and 
$kT_{\rm low}$) and the recombination age ($n_e t$) of the later relaxation phase.

Additionally, the elemental dependence of ionization states is crucial for determining the 
conditions responsible for initially producing the RP.  Fe ions require high temperature 
above 10 keV to be fully ionized, while the lighter elements can reach the full ionization 
with $\sim$1 keV. If Fe ions have higher ionization states than others, that means the RP 
has higher $kT_{\rm high}$.

\subsection{Targets \& Feasibility}
\subsubsection{W49B}

W49B is probably the youngest Galactic mixed-morphology remnant.  While showing both 
RP and enhanced abundances \citep{ozawa09}, it has by far the highest temperature of 
any mixed-morphology remnant.  It is surrounded by dense molecular clouds, which are 
probably responsible for shaping both its morphology and for producing its rapid 
evolution.  It has been classified as both a Type Ia and a core collapse remnant.  It shows 
enhanced metal abundances, and, crucially, is the only mixed-morphology remnant 
showing strong Fe K lines.  W49B is 4 $\times$ 3 arcmin$^2$, and unlike most Galactic 
mixed-morphology SNRs, could be completely mapped using a small number of SXS 
pointings.  
While the metal distribution has been well characterized using {\it Chandra}, the 
degree of over-ionization has not been mapped.

\begin{figure}[htb]
\begin{center}
\includegraphics[width=0.6\textwidth]{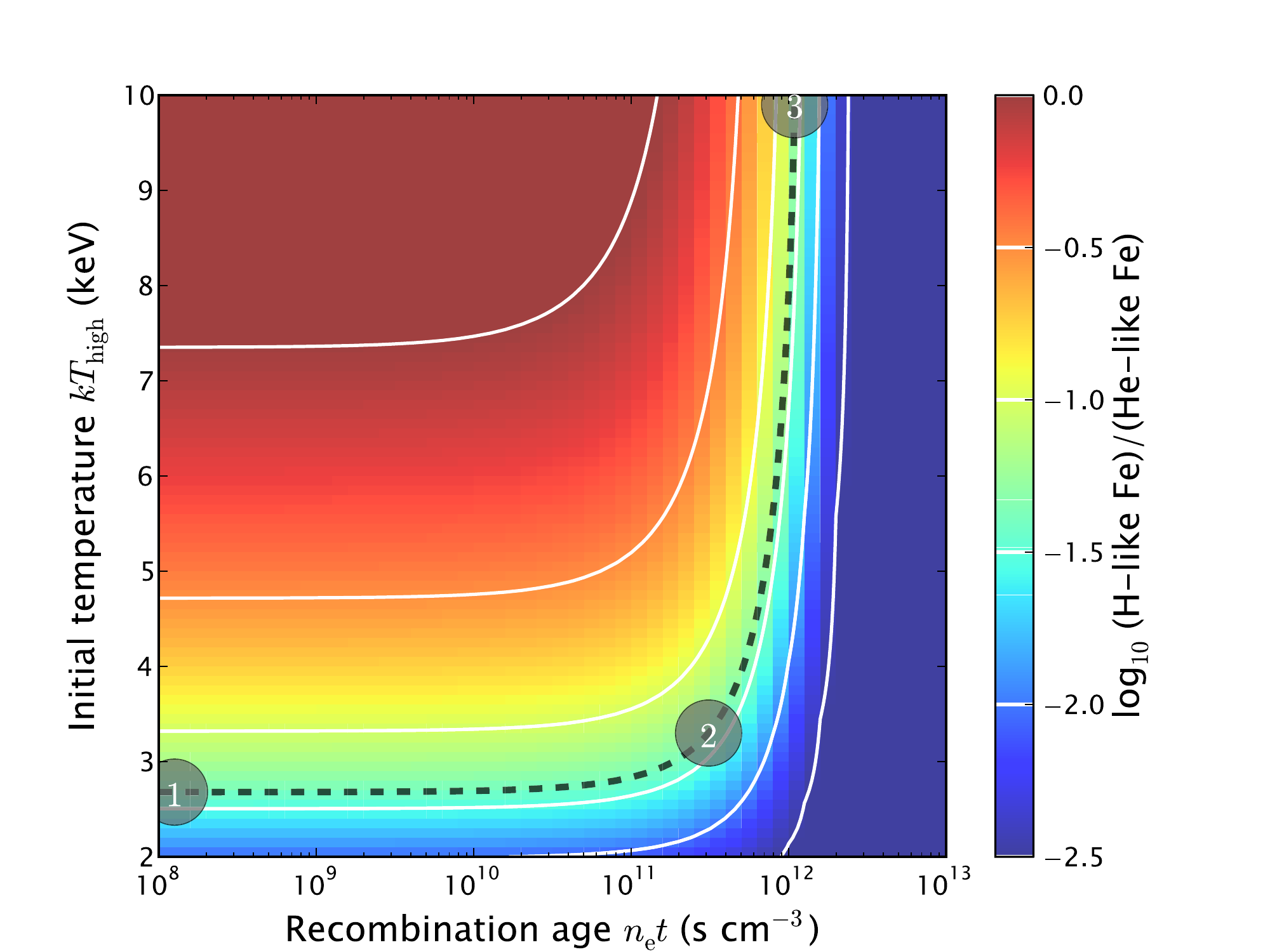}
\caption{Contour map of Fe-K H/He ionic abundance ratio calculated with {\tt SPEX}. 
{\it Suzaku} constrained region for W49B is shown in the dashed line. 
We cannot resolve the initial temperature $kT_{\rm high}$ or the recombination age $n_e t$. 
We have calculated 3 model spectra for W49B with assumptions about the initial 
temperature and recombination age indicated by the 3 large dots.}
\label{fig:w49b_suzaku_contraint}
\end{center}
\end{figure}

As discussed in Section \ref{mixed_morph}, the underlying physical mechanisms causing 
over-ionization in mixed morphology SNRS are a mystery.  We do not know whether the 
plasma in these systems is heating or cooling, or whether the recombination age is short 
or long, since with CCDs one can only resolve the H and He-like lines in the spectra.  
Consequently, we do not know whether the mechanism that leads to cooling is 
rarefaction, conduction, or some other process.  

W49B is the best object for addressing some of these questions. 
The {\it Suzaku} constraints on the initial temperature and recombination age are shown in Figure 
\ref{fig:w49b_suzaku_contraint}.   The initial temperature could range from 2.7 keV to 
higher than 10 keV, and the recombination age could range anywhere from less than 
10$^{8}$ s cm$^{-3}$ to 10$^{12}$ s cm$^{-3}$.

W49B is especially important because of the diagnostic power provided by the Fe-K lines.  
The fine structure of the Fe-K line complex provides the best information about the 
evolutionary stage of RP (as a function $n_e t$).  The longer recombination age results in 
a broader ion population.  The SXS can resolve Fe K$\alpha$ lines into Ly-$\alpha$, He-
like triplets, and satellite lines from lower ionization states. Their flux ratio provides a direct 
measure of the ion population.

\begin{figure}[htb]
\begin{center}
\includegraphics[width=0.49\textwidth]{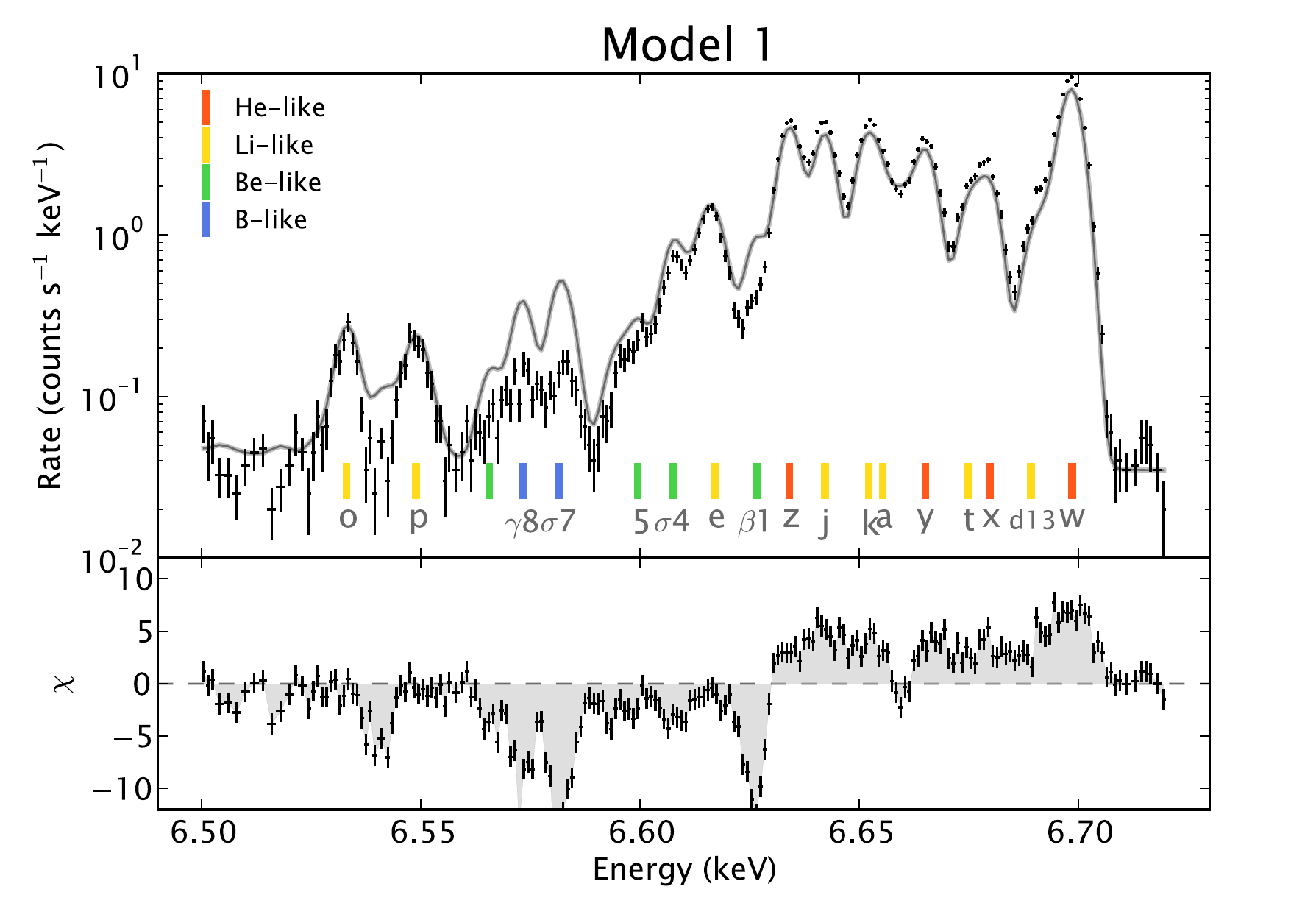}
\includegraphics[width=0.49\textwidth]{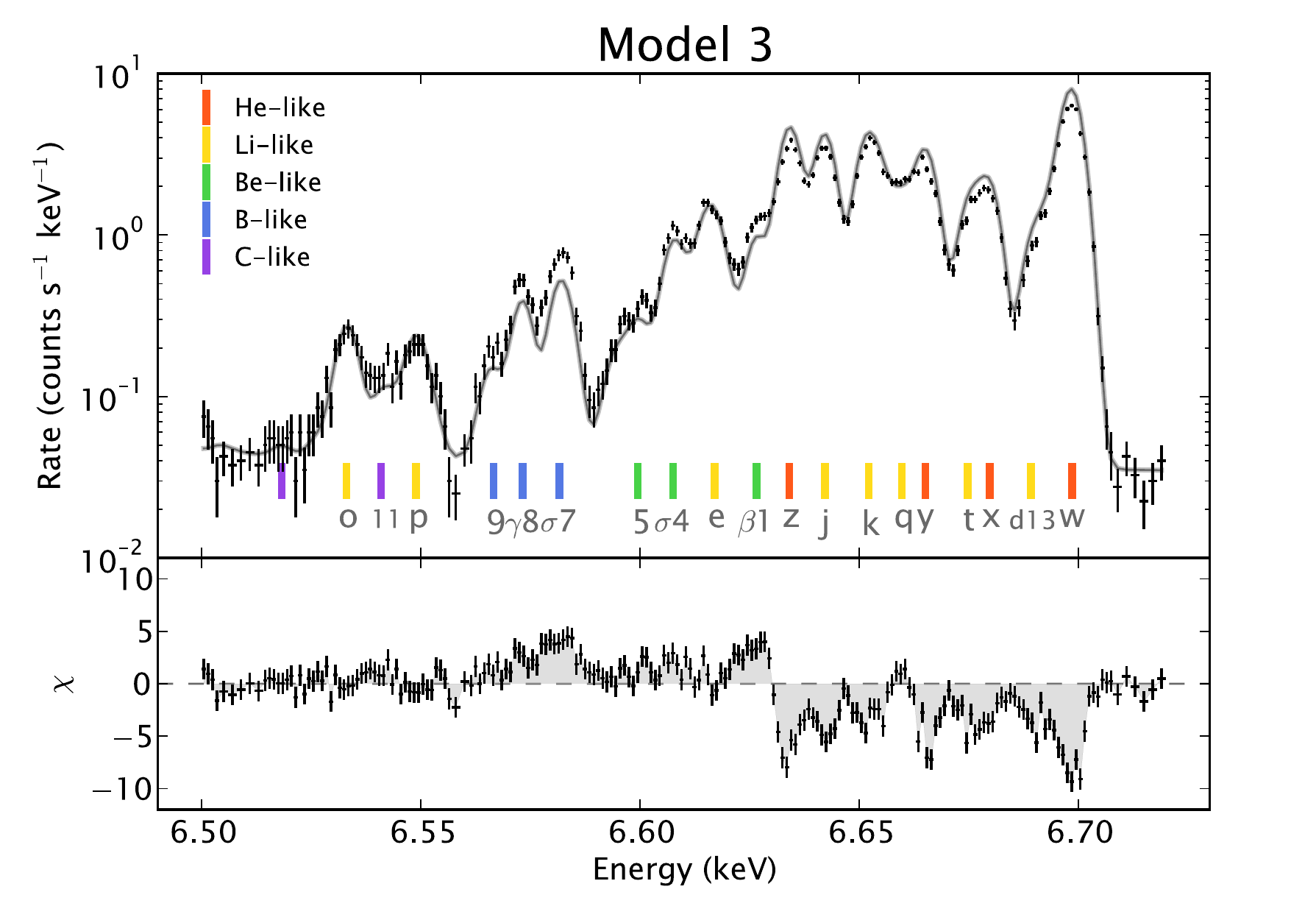}
\caption{Upper panels: 200-ks simulation results of Fe-K$\alpha$ line of W49B 
with {\tt SPEX}, with the assumptions for models 1 and 3 (crosses).  The lower panels 
show the differences between these spectra and that predicted for model 2 (faint curves 
in the upper panels).  The differences in the 3 models are associated with changes in the 
intensities of the satellite lines. }
\label{fig:w49b_models}
\end{center}
\end{figure}

To show the diagnostic capability of {\it ASTRO-H} SXS, we have carried out a set of 
200-ks simulations of the W49B spectrum corresponding to three different assumptions, 
covering the wide range of potential initial temperatures and recombination ages.  Model 
1 has an initial temperature of 2.7 keV and a short recombination age, model 2 has a 
temperature of 3.3 keV and an age of 10$^{11.5}$ s cm$^{-3}$, and model 3 has a 
temperature of 10 keV and an age of 10$^{12}$ s cm$^{-3}$.  As shown in Figure 
\ref{fig:w49b_models}, one very clear way to determine these parameters is from the 
satellite lines, which increase in strength with the recombination age.

\begin{figure}[htb]
\begin{center}
\includegraphics[width=0.5\textwidth]{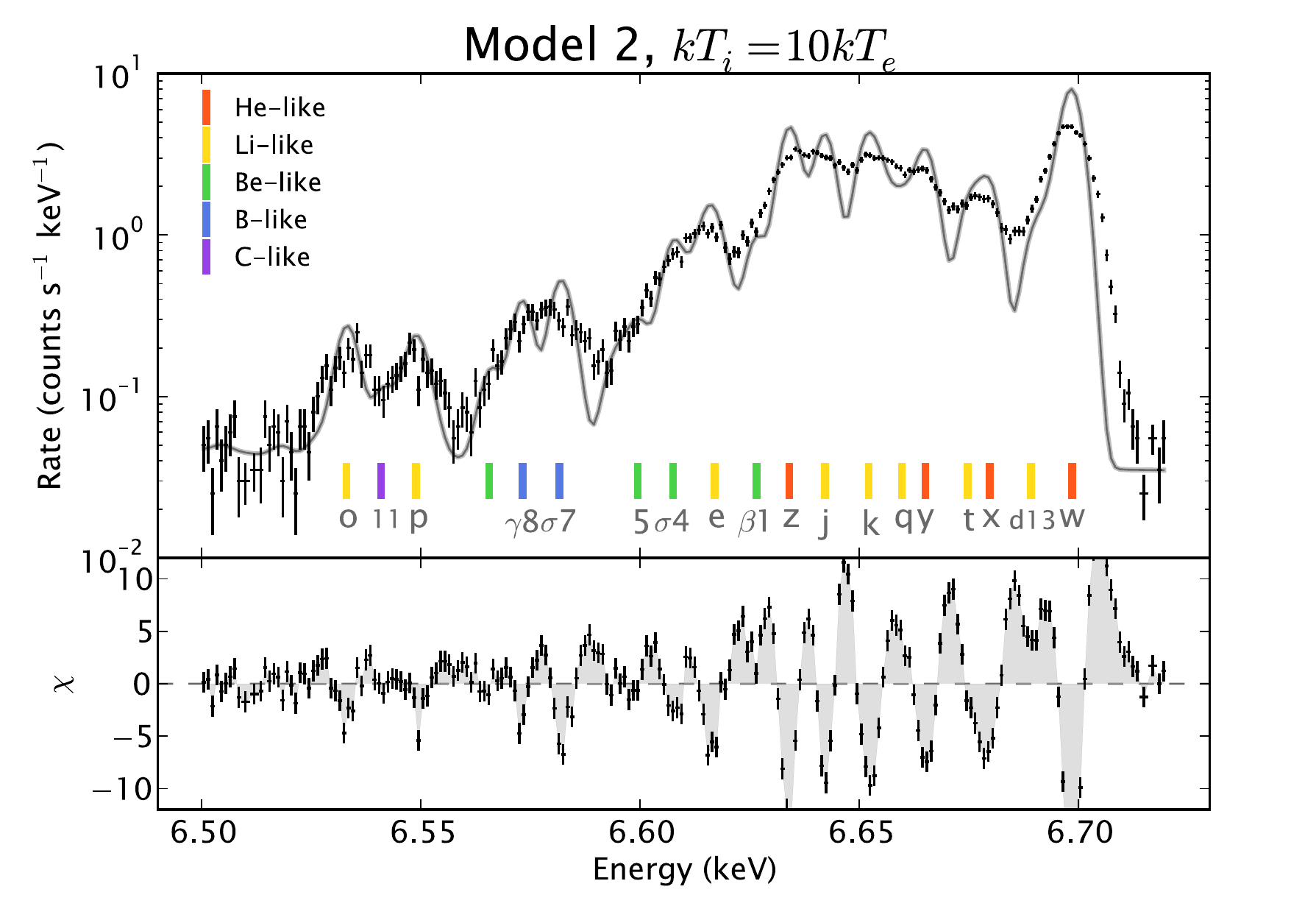}
\caption{Same 200-ks simulation with {\tt SPEX}, but with thermal broadening in case of model 2 (crosses).
We assumed $kT_i=10kT_e$ in the simulation.}
\label{fig:w49b_broadening}
\end{center}
\end{figure}

Similarly, {\it ASTRO-H} SXS observations could potentially distinguish between the main 
two explanations for over-ionization, namely rarefaction or conduction.  If the plasma is 
cooled by rarefaction (adiabatic expansion), then high density and high initial temperature 
are required, which in turn implies ion-electron temperature equilibrium ($kT_i=kT_e$).  
However, if recombining plasmas are produced by conduction, then the process of cooling 
will cause electrons to cool faster than ions, and thus $kT_i>kT_e$.  In this case, 
temperatures measured from thermal broadening will exceed those derived from line 
ratios and continuum emission.  {\it ASTRO-H}, with SXS, will be able to detect this, as 
shown in Figure \ref{fig:w49b_broadening} in the same 200 ks required to measure the 
satellite lines. 

\subsubsection{Other potential targets}

{\bf W44} is the prototypical mixed-morphology SNR.  It shows all the class traits, 
including OH masers, gamma-ray emission, enhanced central abundances, and RP.  It is 
relatively large (35 $\times$ 27 arcmin) so a complete SXS mapping is not possible.  To 
determine the extent of the RP and the distribution of metals, a series of order 5 pointings 
is envisioned, one at the center and one off set in each of the cardinal directions. 

{\bf IC 443} is typically not classified as a mixed-morphology SNR, but shares many 
features with the class.  It is interacting with an H I cloud to its east and with a thin 
molecular cloud along the line of sight.  The highest surface brightness arises in the 
interior, adjacent to the two cloud interaction regions, and there is a hint of a shell in some 
locations around the periphery.  The bright region shows enhanced metal abundances 
and is where the first evidence of over-ionized plasma in any SNR was found using 
{\it ASCA}, and the first strong RRC was found using {\it Suzaku}.   It is possible that IC 443 
represents a remnant in the process of transition into a mixed-morphology remnant. 

{\bf 3C 391}, another prototypical mixed-morphology remnant, shows a clumpy interior 
with a lack of spectral variation (except for a column density gradient).  It shows neither 
evidence for ejecta nor RRC features.  One pointing could thus serve as a "control" for 
other measurements of mixed-morphology remnants, or it could reveal the presence of 
enhanced abundances and/or RP. 

{\bf G65.3$+$5.7} is the nearest mixed-morphology remnant, and the one with the lowest 
column density.  Its very large diameter (4 $\times$ 5 degrees) makes mapping 
impossible. It is the only mixed-morphology remnant whose   forward shock can be 
observed. 

\subsection{Beyond Feasibility}

As evolved SNRs, mixed-morphology remnants are not expected to emit the non-thermal hard X-rays characteristic of relativistic electrons, and hard non-thermal tails are typically not detected from them.  Their age relative to the electron synchrotron cooling time is too old to expect the presence of the TeV electrons responsible for the synchrotron emission seen from the forward shocks of young SNRs.  On the other hand, the detection of $\gamma$-ray emission from many mixed-morphology remnants indicates that acceleration did occur at one time.  The HXI will provide the most sensitive measurements for non-thermal emission above 10 keV for many of these objects, and either stringent limits on acceleration parameters or a surprise detection.

\section{Pulsar Wind Nebulae and their Evolution \label{pwn}}

\subsection{Background and Previous Studies}

While type Ia supernova explosions involving the nuclear detonation of a white dwarf do not lead to the formation of a compact stellar remnant, 
core-collapse explosions of massive ($\geq$8~solar masses) stars give birth to some of the most dense and magnetic stars in the Universe: neutron stars\footnote{The most massive stars are commonly believed to form black holes not discussed in this white paper.}. 
Following their birth, these compact objects powered initially by their fast rotation spew out a relativistic wind of particles and fields (called pulsar wind) which, as it interacts with the surrounding medium, gets shocked leading to the formation of a pulsar wind nebula (PWN).
PWNe have been observed to emit non-thermal radiation across the electromagnetic spectrum from radio to the highest energy gamma-rays. Their emission, often located near the supernova explosion site (at least in the early phases of the SNR evolution), gives the SNR a centrally-filled morphology, thus their other name ``plerions" (for the ancient greek word \textit{pleres}: $\pi \lambda \eta \rho \eta s$).  If the neutron star was given a velocity kick at birth, then, as the SNR ages, the neutron star and associated PWN can move away from the SNR center and eventually interact with and leave the SNR shell.

As the synchrotron bubbles inflated by fast-spinning, highly magnetized, neutron stars,
PWNe provide nearby laboratories to study the physics of relativistic shocks, their interaction with their hosting SNRs, and to search for the missed pulsars in our Galaxy and beyond.
They are also believed to accelerate cosmic rays to TeV energies.
Recent reviews on PWNe can be found in \cite{gaenslerslane06} (on their structure and evolution), \cite{slane08, kp10} (on their high-energy emission), 
and \cite{safi-harb12, safi-harb13} (on recent statistics and the growing class of atypical PWNe).

The most famous and best studied PWN is the Crab Nebula, whose progenitor exploded in AD 1054, and which provided the first confirmation of Baade \& Zwicky's 1934 prediction that neutron stars are born in supernovae
 \citep[see][for a review of its properties]{hester08}. Moving outward, the Crab nebula consists of a 33-ms rotation-powered pulsar, a bright synchrotron nebula observed across the electromagnetic spectrum, bright filamentary ejecta observed in the optical, and a faint freely expanding supernova remnant. Despite deep searches, the outer shock associated with the blast wave of the SNR remains undetected \citep{seward06}, making the Crab nebula a ``pure plerion" or filled-centre SNR.
 Other SNRs however have a composite-type morphology arising from the pulsar-powered (or plerionic) component plus an outer shell representing the location of the blast wave. Such SNRs are referred to in the literature as plerionic Composites.  Out
 of the 314 known SNRs, we know of $\sim$90~SNRs harbouring PWNe \citep{green09, ferrandssh12}.
 
PWNe are initially confined by the supernova ejecta (as in the Crab nebula), but as they evolve their dynamics become affected by the circumstellar or interstellar medium. 
At any stage of PWN evolution, their X-ray emission provides a unique window to probe the freshly injected particles accelerated at the PWN shock and the interaction with the SNR material.
The {\it Chandra} X-ray observatory in particular has provided us with a high-resolution view of these objects, allowing us to study for the first time their fine structure, dynamics, evolution, and pulsar wind properties \citep{kp08}. 

 Even some 45 years after the discovery of the Crab and Vela pulsars in their respective SNRs, we do not know yet why dozens of PWNe lack the shells expected from the supernova explosion (like the Crab) 
 while others are clearly surrounded by SNR shells (like Vela). 
  Are the missing shells associated with a too-low ambient density, and if so, can we infer their progenitors? Is the explosion mechanism different among these seemingly different classes of objects? What role does
  their evolutionary stage (or age) and ambient conditions play in shaping their morphology and multi-wavelength emission? And why do some SNe form PWNe, others magnetars, and still others neutron stars with low magnetic fields?  The lack of observed thermal X-ray emission from these synchrotron-dominated sources remains of the biggest puzzles in this field.

Furthermore, very little is known about their X-ray spectra above $\sim$10~keV. Being hard X-ray sources with spectra characterized by a power-law with a nearly flat photon index, their synchrotron emission
is expected to extend into the hard X-ray band (100's of~keV) with a break that could be associated with synchrotron losses. 
Furthermore,  their 0.5--10 keV X-ray spectra appear to be systematically harder than those of their TeV counterparts \citep{kp10}. Studying the broadband spectrum in the 0.5--600~keV band ({\it ASTRO-H}'s range) 
will fill the gap between the lower energy studies and the gamma-ray studies,
thus shedding light on their magnetic field, ambient density, and acceleration process \citep{reynolds09, gaenslerslane06}.
In particular, gamma-ray studies with H.E.S.S. have opened a new window to study the more evolved objects through the detection of offset PWNe believed to be ``relic" nebulae
resulting from the interaction with the SNR reverse shock, as e.g. for Vela \citep{aharonian06}; see also \cite{blondin01}.
Such multi-wavelength studies are, for the first time, addressing the leptonic versus hadronic models for pulsar winds, shedding light on their composition and other intrinsic properties such as their magnetic field.
 
\subsection{Prospects \& Strategy}
Thanks to the unprecedented sensitivity and spectral resolution of SXS on board {\it ASTRO-H}, we will be able to address the missing SNR shell problem
and search for signatures of the X-ray emitting ejecta in the pure or shell-less plerions. As an example, after accumulating $\sim$500~ks of {\it Chandra} time on the ``Crab-like" PWN G21.5--0.9, long thought to be
a similarly young and a pure plerion, a limb-brightened X-ray feature (SNR blast wave candidate) as well as thermally emitting ``knots" (ejecta candidate)
were found, making G21.5$-$0.9 a textbook example of a young plerionic Composite-type SNR \citep{mathesonssh10, mathesonssh05, bocchino05, ssh01, slane00}.
X-ray studies of a handful other PWNe with {\it Chandra}, {\it XMM-Newton} and {\it Suzaku} have shown evidence for extended thermal X-ray emission, that could be associated with the shock-heated ejecta or the missing SNR shell,
 beyond the known plerions in 3C~58 \citep{gotthelf07, slane04} and G54.1+0.3 \citep{bocchino10}.

Furthermore, the unique broadband coverage of {\it ASTRO-H} in the 0.5--600~keV
will provide broadband spectra of PWNe, filling the gap between soft X-ray spectra (with current missions) and the gamma-ray spectra (with {\it Fermi} in the GeV and H.E.S.S. in the TeV),
allowing us to search for spectral breaks in the hard X-ray band.

PWNe are also highly polarized sources. 
Polarimetric measurements of the brightest objects with the SGD and potentially HXI
will open a new window to study the magnetic fields in PWNe.

\subsection{Targets \& Feasibility}

The relatively small size of PWNe ($\leq$ arcminutes-scale) makes them excellent targets for SXS.
However the relatively poor imaging resolution of SXS and high non-thermal brightness of PWNe (in comparison to any thermal emission associated with the SNR shell or ejecta)
make the detection of any hidden thermal emission challenging. To achieve this latter potentially ground-breaking scientific goal, off-pulsar SXS pointings will have
to be optimized to maximize the detection likelihood of the thermal emission, while minimizing any contamination by the non-thermal emission from the pulsar and PWN.
Here we discuss two young targets that are strong candidates for the study of thermal X-ray emission from PWNe, but that require SXS's spectral
resolution and sensitivity to shed light on the nature of this emission and thus on their supernova progenitors and environment. 

{\bf 3C~58} is an excellent target to advance our understanding in this area.  3C~58
is sufficiently extended that direct contamination from the pulsar can be avoided.
Using {\it XMM-Newton}, \cite{gotthelf07} detected soft ($\sim$0.2 keV) thermal emission
throughout the SNR, and concluded that Ne was likely several times solar. However, with
the limited spectral resolution of {\it XMM-Newton},  it was not possible to separate out other elements or probe 
the dynamical properties of the ejecta.

\begin{figure}[tb]
 \begin{center}
  \includegraphics[width=1.05\textwidth]{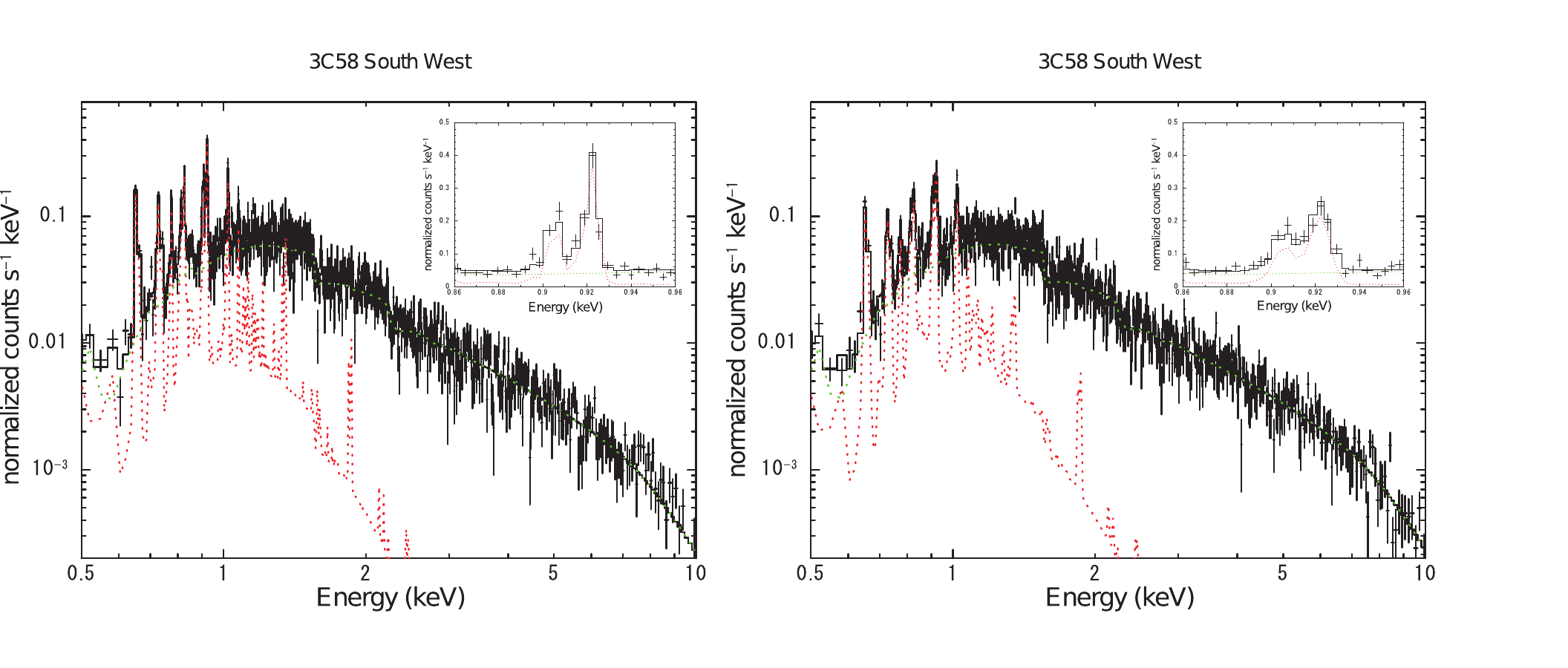} \caption[] {\small
  SXS simulation in the ISM case (left) and the ejecta case (right). Insets show an
  expanded view of Ne lines. }
  \label{fig:3C58} 
 \end{center}
\end{figure}

Figure~\ref{fig:3C58} shows two 100-ks SXS simulations of 3C58 using {\tt simx}. To
begin these simulations, we used the archival {\it XMM-Newton} data and extracted 
a spectrum from the region 2$^{\prime}$.5 south-west from the pulsar, where the thermal flux
is highest while the bright non-thermal central nebula can be avoided
\citep{gotthelf07}. The solar abundance model, except for elevated Ne, provided a
satisfactory fit as previous works. As noted in \cite{gotthelf07}, we stress that this abundance pattern might be
interpreted to be an ISM-origin, not an ejecta-origin, considering recent updates of
`solar-abundance' in which the Ne abundance is systematically increased
\citep{drake05, cunha06}. On the other hand, the {\it XMM-Newton} CCD spectrum also
allowed a fit with the abundance pattern that is calculated from a
18~$M_{\odot}$-progenitor mass case \citep{woosley95}, where Ne/O is slightly enhanced
while Fe/O is depleted compared to the solar ratios. Figure~\ref{fig:3C58} shows
such two different cases, we tentatively call them as the ISM and ejecta cases,
which CCD-type resolution can not distinguish but the SXS can. Contamination of non-thermal emission
from the bright central region is appropriately taken into account. Although
absolute abundances are still difficult to be measured because the thermal continuum is mostly
buried under overwhelming non-thermal emission, the relative abundances between O, Ne,
Fe, and possibly Mg as well as the electron temperature and ionization timescale can be
obtained with a $\sim$10--20\% accuracy (50\% for Mg/O), which allows us to
precisely estimate the progenitor type and the plasma conditions. Furthermore, we note that the line
broadening may be more efficient to discriminate the two cases (see insets of
Figure~\ref{fig:3C58}). Given the young age of 3C58 (suggested to be SN~1181), the
reverse shock velocity in the frame of the ejecta is likely to be expanding with a velocity of
3,000--4,000~km~s$^{-1}$. Combined with almost pure metal composition in the ejecta
case, ion temperatures are still quite high likely to be several hundreds keV
allowing us to measure the line broadening (see also Section 1.3.5).

{\bf G21.5--0.9}:
This PWN is a prototype example of a plerionic Composite-type SNR, originally thought to be a pure plerion like the Crab.
Its age has been estimated as 870$^{+200}_{-150}$~yr \citep{bb08} based on the measured expansion speed of the radio nebula,
while the pulsar's characteristic age is 4.8~kyr \citep{gupta05, camilo06}.
This PWN has been used as a calibration target for {\it Chandra}.
With a flux of a few mCrab in X-rays, it's not too bright to cause pile-up (like the Crab), while also bright enough to
serve as a ``standard" candle.
It has been also used as a cross-calibration study for the instruments onboard currently active observatories: 
{\it Chandra}'s ACIS, {\it Suzaku}'s XIS, {\it Swift}'s XRT, and {\it XMM-Newton}'s EPIC (MOS and pn) for the soft X-ray band, and 
{\it INTEGRAL}'s IBIS-ISGRI, {\it RXTE}'s PCA, and {\it Suzaku}'s HXD-PIN for the hard X-ray band \citep{tsujimoto11}. 
Its size of 5$^{\prime}$-diameter makes it is slightly larger than SXS's field of view (FoV) but fits nicely within the HXI's FoV.

The SXS's unprecedented spectral resolution and sensitivity 
 will help shed light on the nature of the thermal X-ray emission detected from the northern knots, allowing us to differentiate between
a shocked ISM/CSM origin and a shock-heated ejecta origin, as hinted by the {\it Chandra} and {\it XMM-Newton} studies
\citep{mathesonssh10, bocchino05}.  As well, the SXS will allow us to differentiate between a non-thermal or thermal spectrum for the 
eastern limb, which was not possible with the {\it XMM-Newton} and deep {\it Chandra} observations.  
In addition the HXI and SGD will extend the spectrum of the PWN up to as high as 300~keV, allowing for the first time to perform a broadband and 
sensitive spectral study that may reveal a spectral break or at least constrain the power-law photon index across orders of magnitude in energy.

\begin{figure}
 \begin{center}
 {\vspace{-0.5cm}\hspace{-1cm}
\includegraphics[height=2.25in,width=2.25in,angle=0]{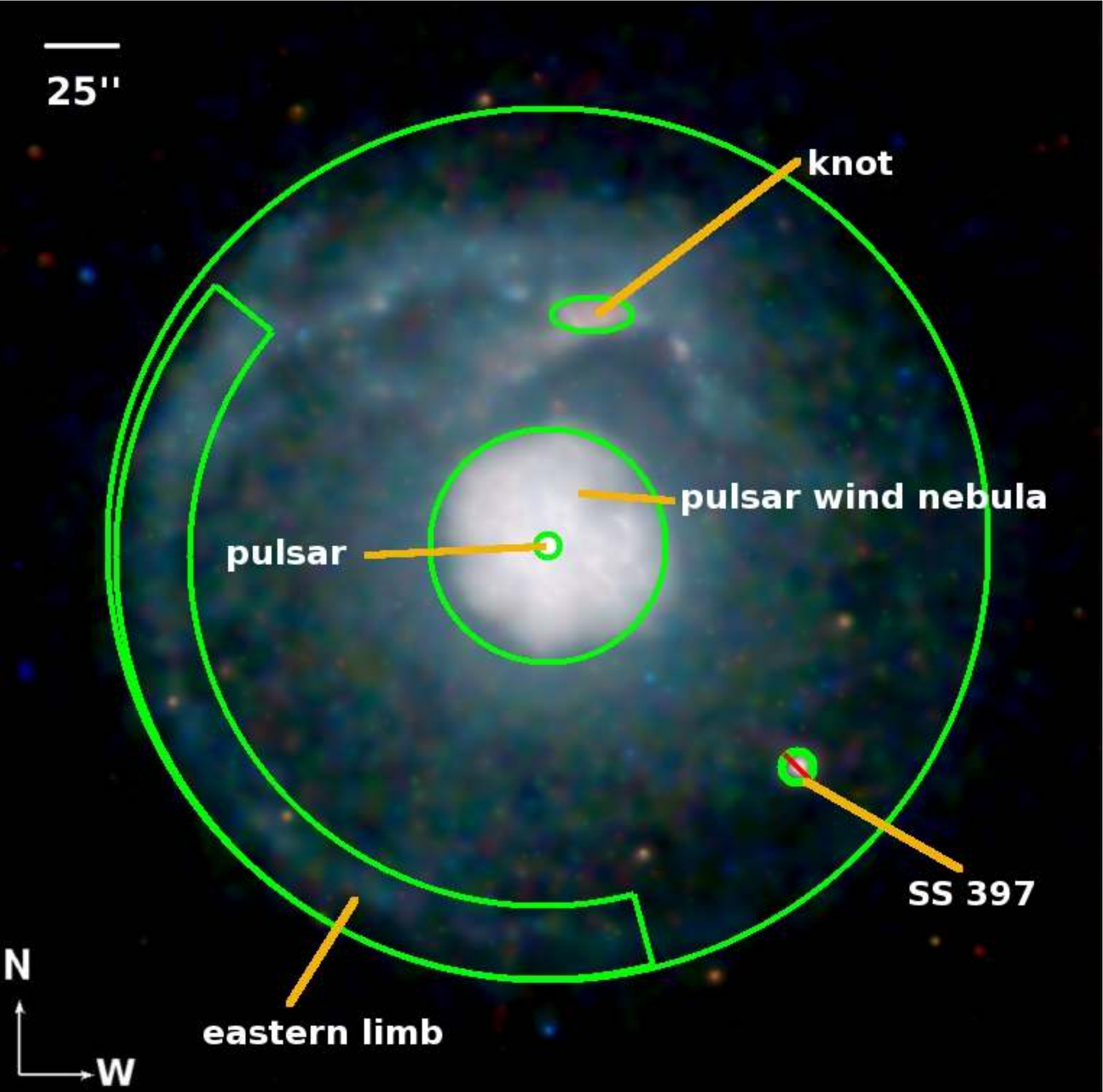}}
{\hspace{+0.5cm}
\includegraphics[height=2.35in,width=2.5in,angle=0]{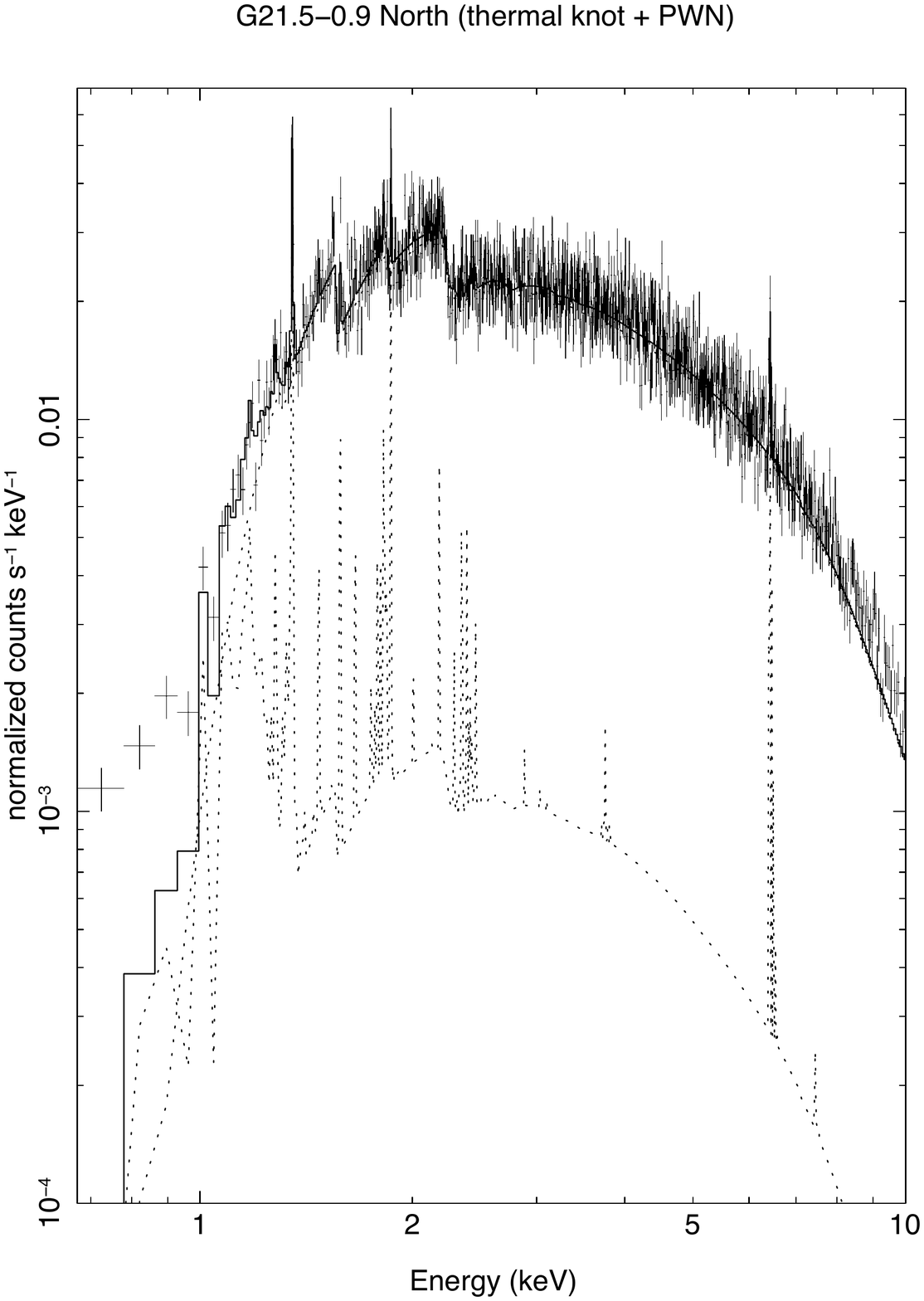}}
\caption{(Left) The {\it Chandra} image of G21.5--0.9 showing the bright central PWN with strong candidate regions
for thermal X-ray emission: the northern knots which likely represent supernova ejecta, 
and the eastern limb showing the long-sought SNR shell \citep{mathesonssh10}.
(Right) A simulated deep (500~ks) SXS spectrum of the thermal emission from the northern knots
for a pointing $\sim$2$^{\prime}$.4 north of the pulsar centred near the bright `knot' shown to the left.
The spectrum was simulated with \texttt{simx} using a thermal
model ($vpshock$) based on the {\it Chandra} data. }
 \label{fig:g21.5ima_knotspec}
\end{center}
\end{figure}

The right panel of Figure~\ref{fig:g21.5ima_knotspec} shows the {\tt sim-x} simulated spectrum of the thermal X-ray emission detected from the knots north of the pulsar/PWN,
acquired with a deep 500~ks SXS simulation pointing $\sim$2$^{\prime}$.4 north of the pulsar (to minimize contamination by the central PWN while
maximizing the thermal emission from the knots). The dotted line illustrates the thermal component ($vpshock$ model with a low ionization timescale of 2$\times$10$^{10}$~cm$^{-3}$~s
and a high temperature of $\sim$5 keV mimicking the {\it Chandra} spectrum) associated with the knots.
Other models were proposed for the emission from this region, including a two-component thermal spectrum with a much lower 
temperature ($kT$$\sim$0.2~keV) plus a non-thermal hard component. Furthermore, while the {\it Chandra} data favour solar abundances (except for S), the {\it XMM-Newton}
data did not rule out enhanced abundances from Mg and Si.

Given the contamination by the central PWN in the SXS pointing at the knot, an additional non-thermal component was accounted for
and was characterized by a power-law model with a photon index of 1.8.
The line emission shown in Figure~\ref{fig:g21.5ima_knotspec} shows the thermal contribution underneath dominant non-thermal PWN emission. 
The SXS pointing and exposure time were chosen such as the lines from the ejecta component will become visible and allow us to constrain the abundance ratios for
Mg, Si, S and Fe, which was not possible with the CCD spectra.
 In particular an Fe-K line becomes clearly visible in the thermal interpretation with a hot temperature and for deep exposures ($\sim$500~ks). 
 Furthermore, as mentioned above for 3C~58, it's likely that the ion temperature is very high and that the
Fe-K line is broadened under the assumption of a similarly young ($\leq$1~kyr) SNR.

Similarly, a deep exposure of the eastern limb will help distinguish between  a power-law and a thermal model (both were acceptable
with the {\it Chandra} data of the eastern limb), and constrain (if thermal in nature) the abundance ratios and the ionization timescale which
is needed to infer the conditions of the ambient medium in which G21.5--0.9 is expanding.

\begin{figure}
\begin{center}
\includegraphics[width=0.65\hsize]{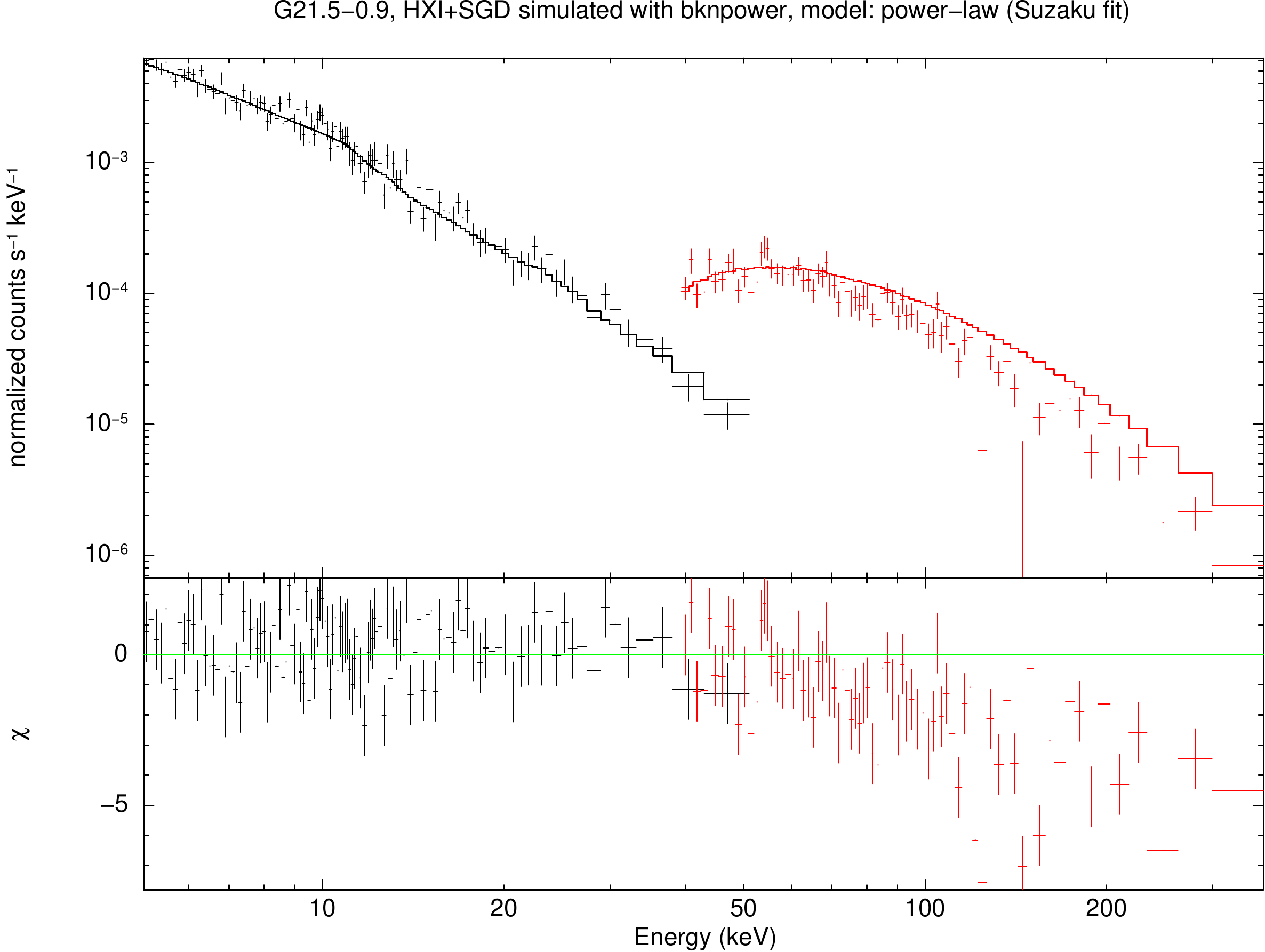}
\caption{A 200~ks HXI (black) and SGD (red) simulated spectrum of the non-thermal, central, PWN  based on the {\it Suzaku} data fitted with a broken power-law model. The model
shown is an absorbed power-law model that provides an adequate fit to {\it Suzaku} data \citep{tsujimoto11}. We note that here only the {\tt nxb} background has been subtracted.}
\label{fig:g21.5hxisgd}
\end{center}
\end{figure}

In Fig.~\ref{fig:g21.5hxisgd}, we show the simulated hard X-ray spectrum of G21.5$-$0.9 with the HXI (5--50 keV) and SGD (40--300 keV).
To simulate the spectrum, we investigated the {\it Suzaku} data which were adequately fitted with a power-law model in the hard band,
as found by \cite{tsujimoto11}. However we also find that a broken power-law with a spectral break within the HXI band can not be ruled out.
Using the XIS and PIN data, the broken power-law model has the following parameters: $\Gamma_1$=1.9 (XIS fit),
$\Gamma_2$=2.3 (HXD-PIN fit) with a spectral break around 35~keV (although poorly constrained). Simulating the HXI and SGD data
with this broken power-law model, a $\geq$200~ks exposure will allow us to distinguish between the broken power-law
and the power-law models. Clearly, the energy coverage will extend beyond {\it NuSTAR}'s hard X-ray band, favouring {\it ASTRO-H} over
{\it NuSTAR} for studying spectral breaks in PWNe.

We acknowledge contributions of B. Guest and H. Matheson (University of Manitoba) towards the simulations shown in Figures 16 and 17.

\subsection{Beyond Feasibility}
Thanks to SXS's sensitivity, {\it ASTRO-H} will potentially open a new window to search for soft thermal X-ray emission that hasn't been detected before
and that may be associated with cold, un-shocked ejecta.
Furthermore, deep X-ray observations of the more evolved PWNe, targeting the PWN regions crushed by the SNR's reverse shock
and observed in the TeV gamma-ray band,  
will address the evolution of PWNe and the nature of their emission from keV to TeV energies.
Finally, polarization studies with HXI and SGD of the brightest PWNe will potentially offer an innovative way to study their magnetic field
driving their dominant non-thermal emission.

\section{Conclusions for Old SNRs and Pulsar Wind Nebulae}


Old SNRs are bright, soft X-ray sources that will have rich spectra dominated by lines in the case of limb-brightened and mixed-morphology remnants, and continuum emission in the case of the pulsar-wind nebulae.  They are astrophysical laboratories for understanding the physics of shocks on the one hand, and for understanding the astrophysical  nature of the ISM and SNRs on the other.  We have highlighted a few of the more important types of observations of the more obvious targets that we expect will be carried out with {\it ASTRO-H}.

\clearpage
\begin{multicols}{2}
\end{multicols}
\end{document}